\documentclass[aps,prd,noeprint,twocolumn,showpacs,amsmath,amssymb]{revtex4-2}
\usepackage{amsmath}
\usepackage{graphicx}
\usepackage{subfigure}
\usepackage{epstopdf}
\usepackage{color}
\usepackage{multirow}
\usepackage{setspace}
\usepackage{overpic}
\usepackage{amssymb}
\usepackage{float}
\usepackage[bookmarksnumbered, pdfstartview=FitH,colorlinks,urlcolor=blue, citecolor=blue,linkcolor=blue]{hyperref}
\usepackage{lineno}
\usepackage{bm}
\usepackage{rotating}
\usepackage[utf8]{inputenc}
\usepackage{enumerate}
\usepackage{mathtools}

\let\oldequation\equation
\let\oldendequation\endequation

\renewenvironment{equation}
  {\linenomathNonumbers\oldequation}
  {\oldendequation\endlinenomath}

\begin{document}

\title{\bf \boldmath
Prospects for $P$ and $C\!P$ violation in $\Lambda_c^+$ decays with polarized beam \\at Super Tau-Charm Facility}
\vspace{4cm}

\author{Hong-Jian Wang$^{1,2}$}
\author{Cheng Wang$^{1,2}$}
\author{Hao Sun$^{3}$}
\author{Pei-Rong Li$^{1,2}$}\email{prli@lzu.edu.cn}
\author{Xiao-Rui Lyu$^{3}$}\email{xiaorui@ucas.ac.cn}
\author{Rong-Gang Ping$^{4,3}$}\email{pingrg@ihep.ac.cn}

\affiliation{$^1$ Lanzhou University, Lanzhou 730000, People’s Republic of China}
\affiliation{$^2$ MOE Frontiers Science Center for Rare Isotopes, Lanzhou University, Lanzhou 730000, People’s Republic of China}
\affiliation{$^3$ University of Chinese Academy of Sciences, Beijing 100049, People’s Republic of China}
\affiliation{$^4$ Institute of High Energy Physics, Beijing 100049, People’s Republic of China}

\date{\it \small \bf \today}

\begin{abstract}
Weak decays of the charmed baryons offer an ideal platform to study parity~($P$) and charge conjugation-parity~($C\!P$) violation in quark sector, to stringently test the Standard Model and search for new physics. 
It is a key goal in the next-generation positron-electron collider, such as the Super Tau-Charm Facility~(STCF). Thanks to the quantum-entangled pair production with super high luminosity and the possibility of beam polarization, STCF provides a unique environment to probe such symmetry violations with unprecedented sensitivity. In this paper,  we evaluate the precisions of the $P$-violating parameters and subsequently obtain the expected sensitivity of the $C\!P$-violating parameters in charmed baryon decays of $\Lambda_{c}^{+}\to p\pi^0$, $p\eta$, $\Lambda K^+$, $\Sigma^0 K^+$, and $\Sigma^+ K^0_S$, regarding to different polarization setups in STCF. The study suggests that the implementation of longitudinal beam polarization in STCF would greatly enhance the experimental capability in studying $P$ and $C\!P$ violation.
\end{abstract}

\maketitle

\oddsidemargin  -0.2cm
\evensidemargin -0.2cm

\section{Introduction}\label{sec:intro}
The Standard Model~(SM) of particle physics is a remarkably successful theoretical framework that describes the properties of elementary particles and their interactions.
However, it falls short of explaining several fundamental questions, such as the origin of the matter-antimatter asymmetry, the source of neutrino masses, and the nature of dark matter and dark energy.
Charge conjugation-parity ($C\!P$) violation in baryons is considered one of the critical factors in addressing the matter-antimatter asymmetry problem~\cite{Sakharov:1967dj}.
To date, experimental observations of $C\!P$ violation have been confirmed in charm and bottom meson systems, notably in $B$ mesons~\cite{BaBar:2001pki} and $D$ mesons~\cite{LHCb:2019hro}.
Recently, LHCb recently reported the first observation of $C\!P$ violation in the bottom baryon sector~\cite{LHCb:2025ray}, marking a significant milestone in the field.
However, it is well known that the magnitude of $C\!P$ violation predicted in the SM is insufficient to account for the observed dominance of matter over antimatter in the universe.
Thus, probing $C\!P$ violation in baryons and searching for potential new physics effects have become crucial to understanding the evolution of the universe.

Parity ($P$) refers to the transformation properties of particles under spatial inversion.
In 1956, Lee and Yang proposed the concept of parity violation observables, specifically the parameter $\alpha$, representing longitudinal polarization of daughter baryon in baryon weak decays~\cite{Lee:1957qs}.
Later that same year, groundbreaking experiment conducted by Wu demonstrated that parity is not conserved in weak decays~\cite{Wu:1957my}, a finding that was subsequently confirmed by numerous experiments.
Nowadays, parity violation has also been widely observed in the decays of charmed baryons~\cite{BESIII:2019odb,BESIII:2022udq,BESIII:2023wrw,BESIII:2024mbf}.
For the decay process $\Lambda_{c}^{+} \to BM$, where $B$ and $M$ represent a baryon with $J^{P}=\frac{1}{2}^{+}$ and a pseudoscalar meson with $J^P=0^-$, respectively, parity-violating observables, $\alpha_{BM}$, $\beta_{BM}$, and $\gamma_{BM}$, are defined in terms of the $s$-wave and $p$-wave decay amplitudes:
\begin{equation}\label{leeyang}
\begin{split}
\alpha_{BM}\coloneqq&2\text{Re}(s^{*}p)/(|s|^{2}+|p|^{2}),\\
\beta_{BM}\coloneqq&2\text{Im}(s^{*}p)/(|s|^{2}+|p|^{2}),\\
\gamma_{BM}\coloneqq&(|s|^{2}-|p|^{2})/(|s|^{2}+|p|^{2}).
\end{split}
\end{equation}
Here, $s$ and $p$ represent the $P$-even and $P$-odd decay amplitudes, respectively.
In a non-relativistic framework, these correspond to the $L = 0$ ($s$-wave) and $L = 1$ ($p$-wave) orbital angular momenta of the baryon-meson system.
If parity is conserved in a given decay process, the polarization parameters $\alpha_{BM}$, $\beta_{BM}$, and $\gamma_{BM}$ are expected to take the values of 0, 0, and $-1$, respectively.
Therefore, polarization measurements serve as a powerful probe for detecting parity violation.
Furthermore, when charge conjugation ($C$) transformation is taken into account, these polarization observables can be further constructed to form $C\!P$-sensitive observables, enhancing the search for $C\!P$ violation in baryon decays.

At present, several observables are employed to investigate $C\!P$ violation in the baryon sector, including the following items:
\begin{enumerate}[(i)]
    \item \textbf{Decay Rate Asymmetry}: The most common approach is to search for $C\!P$ violation by comparing the decay rate ($\Gamma$) with its charge-conjugate counterpart ($\bar{\Gamma}$). This method has been extensively used to study $C\!P$ violation in meson and baryon systems. The $C\!P$-violating observable arising from the decay rate asymmetry is defined as:
    \begin{equation}
        A_{C\!P}^{\Gamma} \coloneqq \frac{\Gamma - \bar{\Gamma}}{\Gamma + \bar{\Gamma}},
    \end{equation}
    where $\Gamma$ and $\bar{\Gamma}$ can be time-dependent or time-integrated.

    \item \textbf{Triple Product Asymmetry}: An alternative method is the measurement of $C\!P$ violation through triple product observables, which have been applied in the four-body decays~\cite{LHCb:2019oke}. The triple product is defined as:
    \begin{equation}
        C_{\hat{T}} \coloneqq \vec{v_1} \cdot (\vec{v_2} \times \vec{v_3}),
    \end{equation}
    where $\vec{v}$ represents observables like momentum in experiments.
    To extract $C\!P$-violating effects, two asymmetry parameters are introduced, based on the event counts $N$:
    \begin{equation}
    \begin{split}
        A_{\hat{T}} &\coloneqq \frac{N(C_{\hat{T}} > 0) - N(C_{\hat{T}} < 0)}{N(C_{\hat{T}} > 0) + N(C_{\hat{T}} < 0)}, \\
        \bar{A}_{\hat{T}} &\coloneqq \frac{\bar{N}(-\bar{C}_{\hat{T}} > 0) - \bar{N}(-\bar{C}_{\hat{T}} < 0)}{\bar{N}(-\bar{C}_{\hat{T}} > 0) + \bar{N}(-\bar{C}_{\hat{T}} < 0)}.
    \end{split}
    \end{equation}
    The corresponding $C\!P$ observable~\cite{Durieux:2015zwa} is then constructed as:
    \begin{equation}
        A_{C\!P}^{\hat{T}} \coloneqq \frac{A_{\hat{T}} - \bar{A}_{\hat{T}}}{2}.
    \end{equation}
    Under $C\!P$ conservation, this observable is expected to be zero.

    \item \textbf{Polarization-based Observables}: The third category focuses on constructing $C\!P$-violating observables through polarization measurements. Under charge conjugation, assuming $C\!P$ symmetry is conserved, the amplitudes of the decay $\bar{\Lambda}_c^- \to \bar{B} \bar{M}$ satisfy the relations $\bar{s} = -s$ and $\bar{p} = p$. This implies that the polarization observables transform as $\bar{\alpha}_{\bar{B}\bar{M}} = -\alpha_{BM}$ and $\bar{\beta}_{\bar{B}\bar{M}} = -\beta_{BM}$. Thus, the $C\!P$-violating observables can be constructed as~\cite{Donoghue:1985ww}:
    \begin{equation}
    \begin{split}
        A_{C\!P}^{\alpha} &\coloneqq \frac{\alpha_{BM} + \bar{\alpha}_{\bar{B}\bar{M}}}{\alpha_{BM} - \bar{\alpha}_{\bar{B}\bar{M}}}, \\
        A_{C\!P}^{\beta} &\coloneqq \frac{\beta_{BM} + \bar{\beta}_{\bar{B}\bar{M}}}{\beta_{BM} - \bar{\beta}_{\bar{B}\bar{M}}}.
    \end{split}
    \end{equation}
Nonzero values for these observables would indicate $C\!P$ violation.
\end{enumerate}

However, several of the above observables have different sensitivities to $C\!P$ violation~\cite{Donoghue:1985ww}:
\begin{equation}\label{sensitivity}
\begin{split}
    A_{C\!P}^{\Gamma} &\propto -\sin\phi' \sin\delta',\\
    A_{C\!P}^{\hat{T}} &\propto \sin\phi \cos\delta,\\
    A_{C\!P}^{\alpha} &\propto -\tan\phi \tan\delta,\\
    A_{C\!P}^{\beta} &\propto \frac{\tan\phi}{\tan\delta},
\end{split}
\end{equation}
where $\phi'$ ($\delta'$) denotes the weak (strong) phase shift between the two interfering amplitudes contributing to the same final states, and $\phi$ ($\delta$) represents the weak (strong) phase shift arising from interference between amplitudes with different partial wave configurations in the same decay process.

These observables have distinct advantages, and the expected precisions of $A_{C\!P}^{\Gamma}$ and $A_{C\!P}^{\hat{T}}$ have been studied in Ref.~\cite{Shi:2019vus}.
It is evident that when the strong phase shift is small, $A_{C\!P}^{\hat{T}}$ and $A_{C\!P}^{\beta}$ are significantly enhanced.
Conversely, when the strong phase shift is large, particularly near $\pi/2$ or $3\pi/2$, $A_{C\!P}^{\alpha}$ becomes substantially amplified.

Clearly, studies on $A_{C\!P}^{\alpha}$ and $A_{C\!P}^{\beta}$ complement each other in searching for $C\!P$ violation across different coverages of the strong phase parameters. To increase the sensitivity power in searching for $C\!P$ violation phenomena, many new approaches have been proposed, such as measuring the comprehensive polarization effect in inclusive decays~\cite{Wang:2022tcm} and studying the decay rate differences in specific phase space region in multi-body decays~\cite{Wang:2024oyi}.
However, no evidence of $C\!P$ violation in charm baryon decays has been found yet.

\begin{table*}[htbp]
	\centering
	\caption{Predicted $A_{C\!P}$ in $\Lambda_{c}^{+}\to p\pi^0$, $\Lambda_{c}^{+}\to p\eta$, $\Lambda_{c}^{+}\to \Lambda K^+$, $\Lambda_{c}^{+}\to \Sigma^0 K^+$, and $\Lambda_{c}^{+}\to \Sigma^+ K^0_S$ from recent theoretical studies in units of $10^{-4}$. In the SU(3) calculatoins in Ref.~\cite{Yang:2025orn}, the superscript $a$ denotes the value is obtained by taking $\tilde{f}_3^{b,c,d}=\tilde{g}_3^{b,c,d}=0$, and $b$ denotes final state re-scattering mechanism.}
	\label{tab:cpv_predictions}
	\begin{tabular}{c c c c c c}
		\hline\hline
		\raisebox{0.2ex}{Predictions} & \raisebox{0.2ex}{$A_{C\!P}^{\alpha_{p\pi^0}}$} & \raisebox{0.2ex}{$A_{C\!P}^{\alpha_{p\eta}}$}  & \raisebox{0.2ex}{$A_{C\!P}^{\alpha_{\Lambda K^{+}}}$} & \raisebox{0.2ex}{$A_{C\!P}^{\alpha_{\Sigma^{0} K^{+}}}$} & \raisebox{0.2ex}{$A_{C\!P}^{\alpha_{\Sigma^{+}K^0_S}}$} \\
        \hline
        He(2024), SU(3)$^{a}$~\cite{He:2024pxh} & $-1.5\pm1.3$ & $\cdots$ & $0.3\pm0.2$ & $\cdots$ & $\cdots$ \\
        He(2024), SU(3)$^{b}$~\cite{He:2024pxh} & $5.5\pm6.4$ & $\cdots$ & $5.0\pm2.3$ & $\cdots$ & $\cdots$ \\
        Cheng(2025), TDA~\cite{Cheng:2025oyr} & $6.97\pm9.17$ & $0.45\pm0.31$ & $6.89\pm0.82$ & $-4.68\pm0.97$ & $-4.68\pm0.97$ \\
        Cheng(2025), IRA~\cite{Cheng:2025oyr} & $6.97\pm9.78$ & $0.45\pm0.31$ & $6.89\pm0.82$ & $-4.68\pm0.98$ & $-4.68\pm0.98$ \\
        He(2025), SU(3)$^{a}$~\cite{Yang:2025orn} & $2.2\pm0.5$ & $\cdots$ & $0.7$ & $0$ & $0$ \\
        He(2025), SU(3)$^{b}$~\cite{Yang:2025orn} & $10.2\pm6.4$ & $\cdots$ & $2.8\pm1.0$ & $-2.1\pm1.5$ & $-2.1\pm1.5$ \\
        \hline\hline
	\end{tabular}
\end{table*}

On the other hand, the choices of the signal processes depend on the reconstructed signal yields and the sizes of the involved phase shifts, which affect the sensitivities in searching for $C\!P$ violation.
A decay with a large strong phase shift is suitable for measuring $A_{C\!P}^{\alpha}$, while a decay with small shift is sensitive for measuring $A_{C\!P}^{\beta}$. 
Recently, a measurement of polarization parameters for $\Lambda_c^+\to\Xi^0 K^+$~\cite{BESIII:2023wrw} showed a significant nonzero strong phase shift~\cite{Wang:2024wrm}.
Based on this work, theorists have estimated strong phase shifts in various processes through SU(3) symmetry~\cite{Geng:2023pkr} and the topological diagram approach~(TDA) combined with the K\"{o}rner-Pati-Woo~(KPW) theorem~\cite{Zhong:2024zme}.
According to these studies~\cite{He:2024pxh,Sun:2024mmk}, five Cabibbo-suppressed~(CS) decays $\Lambda_{c}^{+}\to p\pi^0$, $\Lambda_{c}^{+}\to p\eta$, $\Lambda_{c}^{+}\to \Lambda K^+$, $\Lambda_{c}^{+}\to \Sigma^0 K^+$, and $\Lambda_{c}^{+}\to \Sigma^+ K^0_S$ exhibit relatively large signal yields in experimental reconstruction, along with nonzero weak and strong phase shifts.
Table~\ref{tab:cpv_predictions} lists several recently predicted values of $A_{C\!P}^{\alpha}$ in these charmed baryon decays.
These features make them ideal channels for searching for polarization-induced $C\!P$ violation in next-generation experiments, especially at the Super Tau-Charm Facility~(STCF), which is currently being planned in China~\cite{Luo:2018njj,Cheng:2022tog}.
The STCF is a positron-electron collider operating at center-of-mass energies from 2 to 7~GeV, with an expected data sample approximately 100 times larger than that of the Beijing Electron Positron Collider~(BEPCII)~\cite{Achasov:2023gey,Ye:1987nh,BESIII:2009fln}. 
In addition to high luminosity at STCF, the implementation of beam polarization is expected to enrich the physics program~\cite{Shi:2019vus,Salone:2022lpt} and significantly improve the precision of measurements of $P$ and $C\!P$-violating parameter~\cite{Salone:2022lpt}.

In this paper, we investigate the sensitivities in the measurements of $P$ and $C\!P$-violation observables in the CS decays of $\Lambda_{c}^{+}\to p\pi^0$, $\Lambda_{c}^{+}\to p\eta$, $\Lambda_{c}^{+}\to\Lambda K^+$, $\Lambda_{c}^{+}\to \Sigma^0 K^+$, and $\Lambda_{c}^{+}\to\Sigma^+ K^0_S$ via pair production $e^+e^- \to \Lambda_{c}^{+}\bar{\Lambda}_{c}^{-}$ in the STCF.
By analyzing the Monte Carlo~(MC) samples generated with the \textsc{Fastsim} package in the \textsc{OSCAR} framework~\cite{Ai:2024yqx}, the impact of transverse and longitudinal polarized beams on the experimental accuracy of $P$ and $C\!P$ violation parameters are estimated.
By comparison with the future data collecting plans in the LHCb and Belle II experiments, this paper discusses the prospects of $P$ and $C\!P$ violation in charmed baryons in the STCF.

\section{Methods}\label{sec:methods}
In this section, we describe the theoretical formulism adopted in this study, including the implementation of beam polarization and the statistical methods used for estimating the sensitivities of the physical parameters.
In addition, the expected yields of the $\Lambda_c^+\bar{\Lambda}_c^-$ pairs at STCF and the simulation samples are elaborated.

\subsection{Spin Density Matrix of \texorpdfstring{$e^+e^-\to\Lambda_c^+\bar{\Lambda}_c^-$}{ee2LcLc}}\label{met:sdm}

\begin{figure}[htbp]
	\centering
	\includegraphics[width=\linewidth]{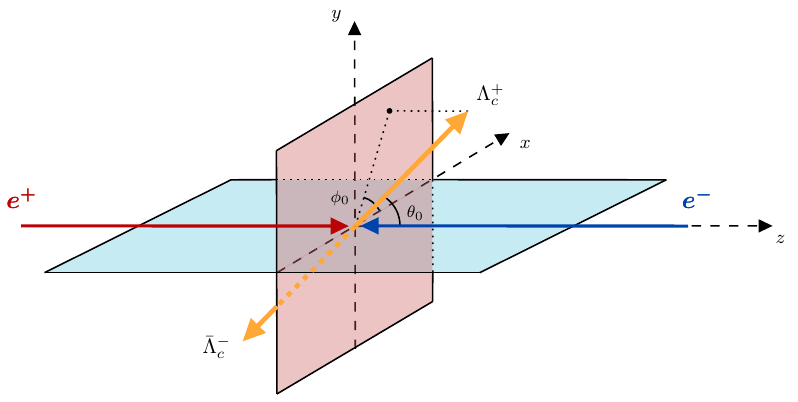}
	\caption{Topological structure of the process $e^{+}e^{-} \to\Lambda_c^+\bar{\Lambda}_c^-$. The $z$-axis is defined along the momentum direction of the positron beam.}
	\label{fig:generation}
\end{figure}

The topological structure of generation is shown in Fig.~\ref{fig:generation}. The momenta of $\Lambda_c^+$ and $\bar{\Lambda}_c^-$ are defined in the center-of-mass~(CM) system of the $e^{+}e^{-}$ pair, and $\theta_{0}$ is the polar angle of $\Lambda_c^+$ in CM system. $\phi_{0}$ is the azimuthal angle of $\Lambda_c^+$ in the CM system, defined as the angle between the projection of its momentum direction onto the $x$-$y$ plane and the positive $x$-axis.

We follow the derivation strategy presented in Refs.~\cite{Cao:2024tvz,Salone:2022lpt}, where the electron mass is neglected and the process is assumed to proceed solely via single-photon exchange.
For the production of $e^{+}e^{-}\to\gamma^*(\lambda_0)\to\Lambda_c^+(\lambda_1)\bar{\Lambda}_c^-(\lambda_2)$, $\lambda_0$, $\lambda_1$, and $\lambda_2$ denote the helicities of the corresponding particles, taking values of $\pm1$ for the virtual photon and $\pm\frac{1}{2}$ for $\Lambda_c^\pm$.
In our treatment, the beam polarizations of the positron and electron are considered separately, and their spin density matrices are expressed in their respective helicity frames as::
\begin{equation}\label{eq:beam}
\begin{aligned}
\rho^- &= \frac{1}{2}
\begin{pmatrix}
1+P_Z & P_T \\
P_T^* & 1-P_Z
\end{pmatrix},~
\rho^+ = \frac{1}{2}
\begin{pmatrix}
1+\bar{P}_Z & \bar{P}_T \\
\bar{P}_T^* & 1-\bar{P}_Z
\end{pmatrix}
\end{aligned}
\end{equation}
where $P_Z$ and $P_T$ are longitudinal and transverse polarization of beam. The spin density matrix element of $\gamma^*$ is given by in the $\Lambda_c^+$ helicity frame:
\begin{equation}\label{rho_gamma}
\rho^{\gamma^*}_{i,j}=\sum_{m,n=\pm1}{\rho^-_{m,n}\rho^+_{-m,-n}D^{1*}_{m,i}(\phi_0,\theta_0,0)D^{1}_{n,j}(\phi_0,\theta_0,0)},
\end{equation}
where the indices $i=\lambda_1-\lambda_2$ and $j=\lambda^{'}_1-\lambda^{'}_2$ correspond to the helicity differences of the initial and final baryon pairs, and $D^{J}_{m,n}(\phi,\theta,\gamma)=e^{-im\phi} \, d^{J}_{m,n}(\theta) \, e^{-in\gamma}$ is Wigner-D function~\cite{grouptheory}.
The density matrix $\rho^{\gamma^*}$ can be written as $\rho^{\gamma^*}=\rho^{\gamma^*}_U+\rho^{\gamma^*}_T+\rho^{\gamma^*}_Z$, where the subscripts $U$, $T$, $Z$ denote the unpolarized, transverse polarization, and longitudinal polarization components, respectively.
The full density matrix $\rho^{\gamma^*}$ is given by:
\begin{widetext}
\begin{equation}\label{rho_gamma_expend}
\scriptsize
\begin{split}
\rho^{\gamma^*}(\phi_0,\theta_0)=\frac{1}{4}&
{\begin{pmatrix}
\displaystyle \frac{1+\cos^2\theta_0}{2}	             & \displaystyle -\frac{\sin\theta_0\cos\theta_0}{\sqrt2} & \displaystyle \frac{\sin^2\theta_0}{2}\\
\displaystyle -\frac{\sin\theta_0\cos\theta_0}{\sqrt2}   & \displaystyle \sin^2\theta_0	                          & \displaystyle \frac{\sin\theta_0\cos\theta_0}{\sqrt2}\\
\displaystyle \frac{\sin^2\theta_0}{2}	                 & \displaystyle \frac{\sin\theta_0\cos\theta_0}{\sqrt2}  & \displaystyle \frac{1+\cos^2\theta_0}{2}
\end{pmatrix}}\\
+\frac{P_T\bar{P}_T}{4}&
{\begin{pmatrix}
\displaystyle \frac{\sin^2\theta_0\cos2\phi_0}{2}	                                      &\displaystyle \frac{\sin\theta_0\cos\theta_0\cos2\phi_0-i\sin\theta_0\sin2\phi_0}{\sqrt2}   &\displaystyle \frac{(1+\cos^2\theta_0)\cos2\phi_0-2i\cos\theta_0\sin2\phi_0}{2}\\
\displaystyle \frac{\sin\theta_0\cos\theta_0\cos2\phi_0+i\sin\theta_0\sin2\phi_0}{\sqrt2} &\displaystyle -\sin^2\theta_0\cos2\phi_0	                                                   &\displaystyle \frac{-\sin\theta_0\cos\theta_0\cos2\phi_0+i\sin\theta_0\sin2\phi_0}{\sqrt2}\\
\displaystyle \frac{(1+\cos^2\theta_0)\cos2\phi_0+2i\cos\theta_0\sin2\phi_0}{2}	          &\displaystyle \frac{-\sin\theta_0\cos\theta_0\cos2\phi_0-i\sin\theta_0\sin2\phi_0}{\sqrt2}  &\displaystyle \frac{\sin^2\theta_0\cos2\phi_0}{2}
\end{pmatrix}}\\
+\frac{1}{4}&
{\begin{pmatrix}
\displaystyle \frac{-2(P_Z+\bar{P}_Z)\cos\theta_0+P_Z\bar{P}_Z(1+\cos^2\theta_0)}{2}          &\displaystyle \frac{(P_Z+\bar{P}_Z)\sin\theta_0-P_Z\bar{P}_Z\sin\theta_0\cos\theta_0}{\sqrt2} &\displaystyle \frac{P_Z\bar{P}_Z\sin^2\theta_0}{2}\\
\displaystyle \frac{(P_Z+\bar{P}_Z)\sin\theta_0-P_Z\bar{P}_Z\sin\theta_0\cos\theta_0}{\sqrt2} &\displaystyle P_Z\bar{P}_Z\sin^2\theta_0                                                      &\displaystyle \frac{(P_Z+\bar{P}_Z)\sin\theta_0+P_Z\bar{P}_Z\sin\theta_0\cos\theta_0}{\sqrt2}\\
\displaystyle \frac{P_Z\bar{P}_Z\sin^2\theta_0}{2}                                            &\displaystyle \frac{(P_Z+\bar{P}_Z)\sin\theta_0+P_Z\bar{P}_Z\sin\theta_0\cos\theta_0}{\sqrt2} &\displaystyle \frac{2(P_Z+\bar{P}_Z)\cos\theta_0+P_Z\bar{P}_Z(1+\cos^2\theta_0)}{2}
\end{pmatrix}}.
\end{split}
\end{equation}
\end{widetext}
For simplicity, under the assumption that $P_T=\bar{P}_T$ due to the Sokolov–Ternov effect~\cite{Sokolov:1963zn}, we define $\hat{P}_T^2 \equiv P_T \bar{P}_T$.
Similarly, we neglect the longitudinal polarization of the positron by setting $\bar{P}_Z = 0$, and redefine $P_Z$ as $\hat{P}_Z$.
After simplification, both $\hat{P}_Z$ and $\hat{P}_T$ reflect single beam polarization. The density matrix reflecting longitudinal beam polarization can be simplified as:
\begin{equation}\label{rho_gamma_expend3}
\frac{1}{4}
\begin{pmatrix}
\displaystyle -\hat{P}_Z\cos\theta_0               &\displaystyle \frac{\hat{P}_Z\sin\theta_0}{\sqrt2} &\displaystyle 0\\
\displaystyle \frac{\hat{P}_Z\sin\theta_0}{\sqrt2} &\displaystyle 0                                    &\displaystyle \frac{\hat{P}_Z\sin\theta_0}{\sqrt2}\\
\displaystyle 0                                    &\displaystyle \frac{\hat{P}_Z\sin\theta_0}{\sqrt2} &\displaystyle \hat{P}_Z\cos\theta_0
\end{pmatrix}.
\end{equation}

The density matrix for the production process is the sum of the contributions of two helicities~\cite{Salone:2022lpt,Cao:2024tvz,Perotti:2018wxm}:
\begin{equation}\label{rho_BB_def}
\rho^{\lambda_1,\lambda_2;\lambda^{'}_1,\lambda^{'}_2}_{B,\bar{B}}=\mathcal{H}_{\lambda_1,\lambda_2}\mathcal{H}^{*}_{\lambda^{'}_1,\lambda^{'}_2}\rho^{\gamma^*}_{\lambda_1-\lambda_2,\lambda^{'}_1-\lambda^{'}_2}.
\end{equation}
Two angular distribution parameters, $\alpha_0$ and $\Delta\Phi$, are defined as $\alpha_0\coloneqq(\mathcal{H}_{\frac{1}{2},\frac{1}{2}}-2\mathcal{H}_{\frac{1}{2},-\frac{1}{2}})/(\mathcal{H}_{\frac{1}{2},\frac{1}{2}}+2\mathcal{H}_{\frac{1}{2},-\frac{1}{2}})$ and $\Delta\Phi\coloneqq\text{Arg}(\mathcal{H}_{\frac{1}{2},-\frac{1}{2}}/\mathcal{H}_{\frac{1}{2},\frac{1}{2}})$, where $\mathcal{H}_{\pm\frac{1}{2},\pm\frac{1}{2}}$ is the helicity amplitudes of the baryon pairs.
The helicity amplitudes of the photon transition to a pair of baryon-antibaryon obey $\mathcal{H}_{-\frac{1}{2},-\frac{1}{2}}=\mathcal{H}_{\frac{1}{2},\frac{1}{2}}$ and $\mathcal{H}_{-\frac{1}{2},+\frac{1}{2}}=\mathcal{H}_{\frac{1}{2},-\frac{1}{2}}$.
Therefore, only two parameters, $\alpha_0$ and $\Delta\Phi$, are sufficient to fully describe the helicity structure of the production amplitude.

The general expression for the joint density matrix of the $B\bar{B}$ pair is
\begin{equation}\label{rho_BB}
\rho_{B\bar{B}}=\sum^3_{\mu,\nu=0}C\sigma^B_\mu\otimes\sigma^{\bar{B}}_\nu,
\end{equation}
where a set of four Pauli matrices $\sigma^B_\mu$~($\sigma^{\bar{B}}_\nu$) in the $B$~($\bar{B}$) rest frame is used and $C_{\mu\nu}$ is a $4\times4$ real matrix representing polarizations and spin correlations of the baryons. 
The matrix $C$ follows the same decomposition into unpolarized, transverse, and longitudinal polarization components as the density matrix $\rho^{\gamma^*}$. The elements of the matrix $C$ are functions of the production angle $\Omega(\theta_0,\phi_0)$ of the $B$ baryon.
The matrix $C_{\mu\nu}$ encapsulates the full spin structure of the baryon-antibaryon system, with its elements determined by both the helicity amplitudes and the beam polarization effects.
The explicit form of $C_{\mu\nu}$ is given by:
\begin{widetext}
\small
\begin{equation}\label{rho_BB_expend}
\begin{split}
C_{\mu\nu}=\frac{3}{2(3+\alpha_0)}&
{\begin{pmatrix}
 1+\alpha_0\cos^2\theta_0                                      & 0                       & \beta_0\sin\theta_0\cos\theta_0 & 0 \\
 0                                                             & \sin^2\theta_0          & 0                                                       & \gamma_0\sin\theta_0\cos\theta_0  \\
 -\beta_0\sin\theta_0\cos\theta_0  & 0 & \alpha_0\sin^2\theta_0  & 0 \\
 0                                                             & -\gamma_0\sin\theta_0\cos\theta_0  &  0 & -\alpha_0-\cos^2\theta_0 
\end{pmatrix}}\\
+\frac{3\hat{P}^2_T}{2(3+\alpha_0)}&
{\begin{pmatrix}
 \alpha_0\sin^2\theta_0 \cos2\phi_0 & -\beta_0\sin\theta_0\sin2\phi_0 & -\beta_0\sin\theta_0\cos\theta_0\cos2\phi_0 & 0 \\
 -\beta_0\sin\theta_0\sin2\phi_0 & (\alpha_0+\cos^2\theta_0)\cos2\phi_0 & -(1+\alpha_0)\cos\theta_0\sin2\phi_0 & -\gamma_0\sin\theta_0\cos\theta_0\cos2\phi_0 \\
 \beta_0\sin\theta_0\cos\theta_0\cos2\phi_0  & (1+\alpha_0)\cos\theta_0\sin2\phi_0 & (1+\alpha_0\cos^2\theta_0)\cos2\phi_0 & -\gamma_0\sin\theta_0\sin2\phi_0 \\
 0 & \gamma_0\sin\theta_0\cos\theta_0\cos2\phi_0 & -\gamma_0\sin\theta_0\sin2\phi_0 & -\sin^2\theta_0\cos2\phi_0
\end{pmatrix}}\\
+\frac{3\hat{P}_Z}{2(3+\alpha_0)}&
{\begin{pmatrix}
 0                         & \gamma_0\sin\theta_0 & 0                   & (1+\alpha_0)\cos\theta_0\\
 \gamma_0\sin\theta_0      & 0                    & 0                   & 0  \\
 0                         & 0                    & 0                   & \beta_0\sin\theta_0 \\
 -(1+\alpha_0)\cos\theta_0 & 0                    & \beta_0\sin\theta_0 & 0
\end{pmatrix}},
\end{split}
\end{equation}
\end{widetext}
with the parameters $\beta_0=\sqrt{1-\alpha^2_0}\sin\Delta\Phi$ and $\gamma_0=\sqrt{1-\alpha^2_0}\cos\Delta\Phi$.

\subsection{Angular Distribution in Cascade Decays}\label{met:formula}
The complete cascade angular distribution for a production process $e^+e^-\to\Lambda^+_c\bar{\Lambda}^-_c$ followed by weak two-body decays of baryon-antibaryon pairs can be obtained following the approach in Ref.~\cite{Perotti:2018wxm}.
To illustrate the derivation, we consider a single-level decay $\Lambda^+_c \to p\eta$ as an example.
When we considering the corresponding charge-conjugated decay $\bar{\Lambda}^-_c\to\bar{p}\eta$ at the same time, the total decay chain is $e^+e^-\to\Lambda^+_c\bar{\Lambda}^-_c\to p\eta\bar{p}\eta$.
Since the $\eta$ is a scalar meson and its angular distribution is trivial .
The resulting cascade angular distribution, denoted as  $\mathcal{P}^{\Lambda^+_c\bar{\Lambda}^-_c}(\boldsymbol{\xi};\boldsymbol{\omega})$ is given by Eq.~\eqref{P_LcLc_level1_expand}, where the coefficient matrix $a_{\mu\nu}^{\bar{\Lambda}^-_c}$ takes the same form as $a_{\mu\nu}^{\Lambda^+_c}$.

The complete angular distribution is characterized by six global parameters, which are collectively denoted by$\boldsymbol{\omega}\coloneqq(\alpha_0, \Delta\Phi, \hat{P}_T, \hat{P}_Z, \alpha_{p\eta}, \bar\alpha_{\bar{p}\eta})$.
The $\boldsymbol{\xi}\coloneqq(\Omega_{\Lambda^+_c}, \Omega_{p}, \Omega_{\bar{p}})$ denotes the full set of kinematic variables that specify a single event in the six-dimensional phase space.
In the cascade decay frame, $\Omega_{\Lambda^+_c}$ is $(\phi_0,\theta_0)$ and has been defined in Section~\ref{met:sdm}.
For the decay $\Lambda^+_c\to p\eta$ and $\bar{\Lambda}^-_c\to \bar{p}\eta$, the kinematic variables of the particles are defined in their respective rest frames, with the coordinate axes constructed starting from the $e^+e^-$ CM frame.
In the rest frame of the $\Lambda_c^+$~($\bar{\Lambda}_c^-$), $\theta_1$~($\theta_1'$) and $\phi_1$~($\phi_1'$) denote the polar and azimuthal angles of the proton~(anti-proton), respectively.

The full angular distribution of the decay is constructed from the spin-correlation matrix $C_{\mu\nu}(\Omega_{\Lambda^+c}; \alpha_0, \Delta\Phi, \hat{P}_T, \hat{P}_Z)$, as given in Eq.~\eqref{P_LcLc_level1_expand}, together with the decay matrices $a_{\mu0}^{\Lambda^+_c}(\Omega_p; \alpha_{p\eta})$ and $a_{\nu0}^{\bar{\Lambda}^-_c}(\Omega_{\bar{p}}; \bar\alpha_{\bar{p}\eta})$.
The decay matrices $a_{\mu\nu}$ represent the transformations of the spin operators (Pauli matrices) $\sigma^{\Lambda^+_c}_\mu$ and $\sigma^{p}_\nu$ defined in the $\Lambda^+_c$ and $p$ baryon helicity frames, respectively~\cite{Perotti:2018wxm}.
They saitisfy $\sigma^{\Lambda^+_c}_\mu=\sum_{\nu=0}^{3}a_{\mu\nu}^{\Lambda^+_c}\sigma^{p}_\nu$.
The explicit form of $a^{\Lambda^+_c}_{\mu\nu}(\Omega_{p};\alpha_{p\eta},\Delta_{p\eta})$, representing the polarization vector transformation of the weak decay~\cite{Lee:1957qs}, is given in Eq.~\eqref{a_expand}.
\begin{widetext}
\begin{equation}\label{P_LcLc_level1_expand}
\mathcal{P}^{\Lambda^+_c\bar{\Lambda}^-_c}(\boldsymbol{\xi};\boldsymbol{\omega})=
\frac{1}{(4\pi)^3}\sum_{\mu,\nu=0}^{3}C_{\mu\nu}(\Omega_{\Lambda^+_c}; \alpha_0, \Delta\Phi, \hat{P}_T, \hat{P}_Z)a_{\mu0}^{\Lambda^+_c}(\Omega_{p}; \alpha_{p\eta})a_{\nu0}^{\bar{\Lambda}^-_c}(\Omega_{\bar{p}}; \bar\alpha_{\bar{p}\eta}).
\end{equation}
\begin{equation}\label{a_expand}
a^{\Lambda^+_c\to p\eta}_{\mu\nu}=\begin{pmatrix}
 1 & 0 & 0 & \alpha_{p\eta} \\
 \alpha_{p\eta}\sin\theta_1\cos\phi_1 & \gamma_{p\eta}\cos\theta_1\cos\phi_1-\beta_{p\eta}\sin\phi_1 &-\beta_{p\eta}\cos\theta_1\cos\phi_1-\gamma_{p\eta}\sin\phi_1 & \sin\theta_1\cos\phi_1 \\
 \alpha_{p\eta}\sin\theta_1\sin\phi_1& \beta_{p\eta}\cos\phi_1+\gamma_{p\eta}\cos\theta_1\sin\phi_1 & \gamma_{p\eta}\cos\phi_1-\beta_{p\eta}\cos\theta_1\sin\phi_1& \sin\theta_1\sin\phi_1 \\
 \alpha_{p\eta}\cos\theta_1 &-\gamma_{p\eta}\sin\theta_1 & \beta_{p\eta}\sin\theta_1 & \cos\theta_1
\end{pmatrix}.
\end{equation}
\end{widetext}
In Eq.~\eqref{a_expand}, the parameters $\beta_{p\eta}$ and $\gamma_{p\eta}$ are defined as follows:
\begin{equation}\label{betagamma}
\begin{split}
\beta_{p\eta}&=\sqrt{1-\alpha^2_{p\eta}}\sin\Delta_{p\eta},\\
\gamma_{p\eta}&=\sqrt{1-\alpha^2_{p\eta}}\cos\Delta_{p\eta},
\end{split}
\end{equation}
implying that $\alpha^2_{p\eta}+\beta^2_{p\eta}+\gamma^2_{p\eta}=1$. For the single-level decay, only the first column $a_{\mu0}(\Omega_{p};\alpha_{p\eta})$ is used and it depends only on the decay parameter $\alpha_{p\eta}$. 
For the double-level decay such as $e^+e^-\to\Lambda^+_c\bar{\Lambda}^-_c$ with $\Lambda^+_c\to\Lambda K^+$, $\Lambda\to p\pi^-$, and its charge-conjugated process, $\bar{\Lambda}^-_c\to\bar\Lambda K^-$, $\bar\Lambda\to \bar p\pi^+$, the cascade angular distribution formula is similar with the \eqref{P_LcLc_level1_expand}.
Detailed derivations can be found in the Ref.~\cite{Salone:2022lpt}. 

Estimating the uncertainties for simultaneous decays of both $\Lambda_c^+$ and $\bar{\Lambda}_c^-$ is challenging due to the low detection efficiency and small branching fraction in $\Lambda^+_c$ decay.
To better reflect realistic experimental conditions, a single-tag (ST) method is employed in the estimation.
The ST angular distributions are obtained by integrating over the unmeasured kinematic variables.
As an example, the ST angular distribution for the single-level decay of the $\Lambda^+_c$ reduces to the following form fromEq.~\eqref{P_LcLc_level1_expand}:
\begin{equation}\label{P_Lc_level1_expand}
  \begin{split}
    \mathcal{P}^{\Lambda^+_c}(\boldsymbol{\xi};\boldsymbol{\omega})=\frac{1}{(4\pi)^2}\sum_{\mu=0}^{3}C_{\mu0}\cdot a_{\mu0}^{\Lambda^+_c},
  \end{split}
\end{equation}
where $\boldsymbol{\xi}:=(\Omega_{\Lambda^+_c},\Omega_p)$. 
For the double-level decay process $e^+e^-\to\Lambda^+_c\bar{\Lambda}^-_c$ with subsequent decays $\Lambda^+_c\to\Lambda K^+$ and $\Lambda\to p\pi^-$, the ST angular distribution can be expressed as:
\begin{equation}\label{P_Lc_level2_expand}
  \begin{split}
    \mathcal{P}^{\Lambda^+_c}(\boldsymbol{\xi};\boldsymbol{\omega})=\frac{1}{(4\pi)^3}\sum_{\mu=0}^{3}C_{\mu0}\cdot\sum_{\mu'=0}^3 a_{\mu\mu'}^{\Lambda^+_c}a_{\mu'0}^{\Lambda}.
  \end{split}
\end{equation}
The decay channel $\Lambda_{c}^{+}\to\Sigma^+ K^0_S$, followed by $\Sigma^+\to p\pi^0$, shares the same angular distribution formula as the channel $\Lambda_{c}^{+}\to\Lambda K^+$ with $\Lambda\to p\pi^-$.
For the triple-level decay process $e^+e^-\to\Lambda^+_c\bar{\Lambda}^-_c$ with the decay chain $\Lambda^+_c\to\Sigma^0 K^+$, $\Sigma^0\to\gamma\Lambda$, and $\Lambda\to p\pi^-$ in the ST analysis, we consider that the electromagnetic decay $\Sigma^0\to\gamma\Lambda$ conserves parity.
Therefore, the non-zero elements of the decay matrix $a^{\Sigma^0}_{\mu\nu}$ can be expressed as~\cite{Wang:2016elx}:
\begin{equation}\label{P_em_nonzero}
  \begin{split}
    a^{\Sigma^0}_{0,0}&=1,\\
    a^{\Sigma^0}_{1,3}&=-\sin\theta_2\cos\phi_2,\\
    a^{\Sigma^0}_{2,3}&=-\sin\theta_2\sin\phi_2,\\
    a^{\Sigma^0}_{3,3}&=-\cos\theta_2,
  \end{split}
\end{equation}
where $\theta_2$ and $\phi_2$ are the polar and azimuthal angles of $\Lambda$ in $\Sigma^0$ rest frame, respectively.

\subsection{Monte Carlo Simulation and Event Selection}\label{met:mc}
Monte Carlo simulation for this analysis is performed using the fast simulation framework within the \textsc{OSCAR} software, developed for the STCF.
\textsc{OSCAR} is built on the \textsc{SNiPER} architecture, utilizing \textsc{DD4hep} for detailed detector description and \textsc{PODIO} for data modeling, providing a modular and efficient platform for large-scale MC sample production~\cite{Huang:2022bkz}.
The fast simulation balances computational efficiency and accuracy by parametrizing detector responses instead of full \textsc{Geant4}-based tracking, significantly accelerating event processing while preserving key detector effects.

Event generation is conducted at a center-of-mass energy of 4.68~GeV, chosen due to the notable transverse polarization of $\Lambda_c^{+}$ baryons produced at this energy, which enhances sensitivity to polarization and $C\!P$-violation parameters.
Signal events for the process $e^{+}e^{-} \to \Lambda_c^{+} \bar{\Lambda}_c^{-}$ are generated with $\Lambda_c^{+}$ decaying exclusively to the studied final state, and $\bar{\Lambda}_c^{-}$ decaying inclusively. 
Physics processes such as initial state radiation (ISR) and beam energy spread are simulated using the \textsc{KKMC} generator~\cite{Jadach:2000ir,Jadach:1999vf}.
Decay modeling is handled by \textsc{EvtGen}~\cite{Lange:2001uf,Ping:2008zz} with branching fractions from the latest PDG compilation~\cite{ParticleDataGroup:2024cfk}, supplemented by the LundCharm model~\cite{Chen:2000tv,Yang:2014vra} for unmeasured decay modes.
Final state radiation (FSR) is implemented using \textsc{PHOTOS}~\cite{Barberio:1990ms}.

Two kinds of MC samples are generated.
One is a phase space~(PHSP) sample used for MC integration to estimate the efficiency, and the other is a pseudo-data sample generated with the amplitude model, used to evaluate the $C\!P$ sensitivity.
Input parameters for different processes are listed in Table~\ref{para}.
It should be noted that $C\!P$ conservation is assumed in this study.
This is justified by the fact that $C\!P$ violation in charmed baryons is expected to be very small, typically at the level of $10^{-4}$–$10^{-5}$, which is well below the sensitivity of our measurement.
Therefore, using $C\!P$-conserving samples to estimate the precision of $C\!P$-violating observables is a valid approach, introducing a relative uncertainty of approximately 0.3\% in the precision estimation.

\begin{table}[htbp]
    \small
	\centering
	\caption{Asymmetry parameters employed in the simulation.}
	\label{para}
	\begin{tabular}{c c c c}
		\hline\hline
		Processes                                                 & Parameters                        & Values           & Sources   \\
		\hline
		\multirow{2}*{$e^+e^-\to\Lambda^+_c\bar{\Lambda}^-_c$}    & $\alpha_{0}$~($\bar{\alpha}_{0}$) & $ 0.10$          & Estimated \\
		  ~                                                     & $\Delta\Phi$~($\bar{\Delta}\Phi$) & $-0.50$          & Estimated \\
        \hline
		$\Lambda^+_c\to p\pi^0$                                   & $\alpha_{p\pi^0}$                 & $-0.11$            & Ref.~\cite{Zhong:2024qqs} \\
        $\bar{\Lambda}^-_c\to\bar{p}\pi^0$                        & $\alpha_{\bar{p}\pi^0}$           & $-\alpha_{p\pi^0}$ & $C\!P$ conservation \\
        \hline
		$\Lambda^+_c\to p\eta$                                    & $\alpha_{p\eta}$                  & $-0.42$           & Ref.~\cite{Geng:2023pkr} \\
        $\bar{\Lambda}^-_c\to\bar{p}\eta$                         & $\alpha_{\bar{p}\eta}$            & $-\alpha_{p\eta}$ & $C\!P$ conservation \\
        \hline
		\multirow{2}*{$\Lambda^+_c\to\Lambda K^+$}                & $\alpha_{\Lambda K^+}$            & $-0.52$           & LHCb~(2024)~\cite{LHCb:2024tnq} \\
        ~                                                         & $\Delta_{\Lambda K^+}$            & $ 2.74$           & LHCb~(2024)~\cite{LHCb:2024tnq} \\
        \multirow{2}*{$\bar{\Lambda}^-_c\to\bar{\Lambda} K^-$}    & $\alpha_{\bar{\Lambda} K^-}$      & $-\alpha_{\Lambda K^+}$  & $C\!P$ conservation \\
        ~                                                         & $\Delta_{\bar{\Lambda} K^-}$      & $-\Delta_{\Lambda K^+}$  & $C\!P$ conservation \\
        \hline
		\multirow{2}*{$\Lambda^+_c\to\Sigma^0 K^+$}               & $\alpha_{\Sigma^0 K^+}$           & $-0.52$            & Ref.~\cite{Geng:2023pkr} \\
  	~                                                         & $\Delta_{\Sigma^0 K^+}$           & $-0.59$            & Ref.~\cite{Geng:2023pkr} \\
        \multirow{2}*{$\bar{\Lambda}^-_c\to\bar{\Sigma}^0 K^-$}   & $\alpha_{\bar{\Sigma}^0 K^-}$     & $-\alpha_{\Sigma^0 K^+}$ & $C\!P$ conservation \\
        ~                                                         & $\Delta_{\bar{\Sigma}^0 K^-}$     & $-\Delta_{\Sigma^0 K^+}$ & $C\!P$ conservation \\
		\hline
		\multirow{2}*{$\Lambda^+_c\to\Sigma^+ K^0_S$}             & $\alpha_{\Sigma^+ K^0_S}$         & $-0.52$           & Ref.~\cite{Geng:2023pkr} \\
  	~                                                         & $\Delta_{\Sigma^+ K^0_S}$         & $-0.59$           & Ref.~\cite{Geng:2023pkr} \\
        \multirow{2}*{$\bar{\Lambda}^-_c\to\bar{\Sigma}^+ K^0_S$} & $\alpha_{\bar{\Sigma}^- K^0_S}$   & $-\alpha_{\Sigma^+ K^0_S}$ & $C\!P$ conservation \\
        ~                                                         & $\Delta_{\bar{\Sigma}^- K^0_S}$   & $-\Delta_{\Sigma^+ K^0_S}$ & $C\!P$ conservation \\
        \hline
		$\Lambda\to p\pi^-$                                       & $\alpha_{p\pi^-}$                 & $0.747$  & PDG(2024)~\cite{ParticleDataGroup:2024cfk} \\
        $\bar{\Lambda}\to\bar{p}\pi^+$                            & $\alpha_{\bar{p}\pi^+}$           & $-0.757$ & PDG(2024)~\cite{ParticleDataGroup:2024cfk} \\
        \hline
		$\Sigma^+\to p\pi^0$                                      & $\alpha_{p\pi^0}$                 & $-0.982$  & PDG(2024)~\cite{ParticleDataGroup:2024cfk} \\
        $\bar{\Sigma}^-\to\bar{p}\pi^0$                           & $\alpha_{\bar{p}\pi^0}$           & $ 0.990$ & PDG(2024)~\cite{ParticleDataGroup:2024cfk} \\
		\hline\hline
	\end{tabular}
\end{table}

A single-tag method is adopted, and all final-state particles are explicitly reconstructed.
Four decay modes of the $\Lambda_c^+$ baryon are studied, each involving intermediate states that require dedicated reconstruction. 
For $\Lambda^+_c \to p\pi^0$ and $\Lambda^+_c \to p\eta$, the $\pi^0(\eta)$ meson is reconstructed from $\gamma\gamma$; for $\Lambda^+_c \to \Lambda K^+$, the $\Lambda$ is reconstructed via $p\pi^-$; for $\Lambda^+_c \to \Sigma^0 K^+$, the $\Sigma^0$ is reconstructed through $\gamma\Lambda$, with $\Lambda \to p\pi^-$; and for $\Lambda^+_c \to \Sigma^+ K^0_S$, the $\Sigma^+$ is reconstructed through $p\pi^0$ with $\pi^0\to\gamma\gamma$, while $K^0_S$ is reconstructed via $\pi^+\pi^-$.

Charged tracks are required to fall within the acceptance region $|\cos\theta|<0.93$, where $\theta$ is the polar angle defined relative to the symmetry axis of the main drift charmber~(the z-axis).
Additionally, tracks not originating from a $K^0_S$ meson or a $\Lambda$ baryon are required to have a distance of closest approach to the interaction point (IP) satisfying $|r_z|<10$~cm and $|r_{xy}|<1$~cm, where $r_z$ and $r_{xy}$ denote the distances along the beam axis and in the transverse plane, respectively.
For the proton and pion that come from $K^0_S$ or $\Lambda$ decays, a requirement of $|r_z|<20$~cm is applied, while no constraint is imposed on $|r_{xy}|$.
Particle identification (PID) is performed by combining information from ionization energy loss ($dE/dx$) and time-of-flight (TOF) measurements. For each charged track, likelihoods $\mathcal{L}(h)$ are calculated under different mass hypotheses $h = p, K, \pi$, and the particle type is assigned based on likelihood comparisons.
Proton candidates must satisfy $\mathcal{L}(p)>0$ and have a higher probability than both kaon and pion hypotheses; kaons are required to satisfy $\mathcal{L}(K) > 0$ and $\mathcal{L}(K)>\mathcal{L}(\pi)$.
In the reconstruction of $\Lambda$ and $K^0_S$ decays, no PID requirement is imposed on the daughter pions.
Showers are selected with a minimum energy of 25~MeV in the barrel region ($|\cos\theta|<0.8325$) or 50~MeV in the endcap ($0.8325<|\cos\theta|<0.9445$), and must be separated from any charged track by at least $10^\circ$.
The $K^0_S\to\pi^+\pi^-$ and $\Lambda\to p\pi^-$ candidates are required to satisfy a primary vertex fit with $\chi^2 < 100$, decay length over error $L/\sigma_L > 2$, and invariant mass windows of $(0.487,0.511)~\text{GeV}/c^2$ and $(1.111,1.121)~\text{GeV}/c^2$, respectively.
The invariant mass of $\Sigma^0\to\gamma\Lambda$ is required to lie within $(1.179,1.203)~\text{GeV}/c^2$, and for $\Sigma^+\to p\pi^0$, within $(1.176,1.200)~\text{GeV}/c^2$. The $\pi^0\to\gamma\gamma$ candidates must pass a kinematic fit with $\chi^2 < 20$, and their invariant mass must lie within $(0.115,0.150)~\text{GeV}/c^2$. For $\eta\to\gamma\gamma$, a similar kinematic fit is applied with $\chi^2<20$, and the invariant mass is required to be within the window $(0.510,0.570)~\text{GeV}/c^2$.
Two variables are used in the final selection: the energy difference $\Delta E$ and the beam-constrained mass $M_{\text{BC}}$.
The energy difference is defined as $\Delta E = E_{\Lambda_c^+}-E_{\text{beam}}$, where $E_{\Lambda_c^+}$ is the reconstructed energy of the $\Lambda_c^+$ candidate and $E_{\text{beam}}$ is the beam energy in the center-of-mass frame.
The beam-constrained mass is defined as $M_{\text{BC}} = \sqrt{E_{\text{beam}}^2 - |\vec{p}_{\Lambda_c^+}|^2}$, where $\vec{p}_{\Lambda_c^+}$ is the reconstructed momentum of the $\Lambda_c^+$ candidate in the $e^+e^-$ CM system.
The $\Delta E$ requirement is mode-dependent: for $\Lambda_c^+\to p\pi^0$ and $\Lambda_c^+\to p\eta$, a broader window of $\Delta E\in(-70,70)~\text{MeV}$ is used to accommodate the poorer energy resolution from $\pi^0(\eta)\to\gamma\gamma$; for the other three modes, $\Lambda_c^+\to \Lambda K^+$, $\Sigma^0 K^+$, and $\Sigma^+ K^0_S$, a tighter requirement of $\Delta E\in(-25,20)~\text{MeV}$ is applied. The $M_{\text{BC}}$ signal window $M_{\text{BC}}\in(2.282,2.291)~\text{GeV}/c^2$ is selected as signal region for all four decay modes under study. The four-momenta used to calculate $\boldsymbol{\xi}_{i}$ are refined via a kinematic fit imposing mass constraints on all resonances according to their PDG values~\cite{ParticleDataGroup:2024cfk}.
Overall, since background contributions are not taken into account in this analysis, the estimated efficiency may be somewhat optimistic. Nevertheless, this simplification does not affect the final conclusions of the study.

\begin{figure*}[htbp]
	\centering
	\includegraphics[width=0.9\linewidth]{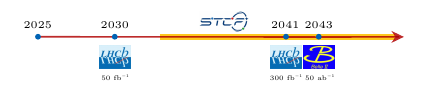}
	\caption{Data taking plan of different experiments in high luminosity with similar physical goals.}
	\label{fig:timeline}
\end{figure*}

\subsection{Fit Strategy}\label{met:fit}
The likelihood function is constructed from the joint probability density function (PDF) by
\begin{equation}
    \label{likelihood}
    \mathcal{L}_{\text{total}}=\prod_{i=1}^{N_{\text{data}}}f_{s}(\boldsymbol{\xi}_{i};\boldsymbol{\omega}).
\end{equation}
Here, $f_{s}(\boldsymbol{\xi}_{i};\boldsymbol{\omega})$ denotes the PDF of the signal process, $N_{\text{data}}$ is the total number of events in the data, and $i$ is the index of each event. The signal PDF $f_{s}(\boldsymbol{\xi}_{i};\boldsymbol{\omega})$ is formulated as
\begin{equation}
	\label{signalPDF}
	f_{s}(\boldsymbol{\xi}_{i};\boldsymbol{\omega})=\frac{\epsilon(\boldsymbol{\xi}_{i})|\mathcal{P}^{\Lambda^+_c}(\boldsymbol{\xi}_{i};\boldsymbol{\omega})|^{2}}{\int\epsilon(\boldsymbol{\xi}_{i})|\mathcal{P}^{\Lambda^+_c}(\boldsymbol{\xi}_{i};\boldsymbol{\omega})|^{2}\text{d}\boldsymbol{\xi}_{i}}~,
\end{equation}
where $\boldsymbol{\xi}_{i}$ represents the set of kinematic angular observables ($\theta_{0,1,2,3}$ and $\phi_{1,2}$), and $\boldsymbol{\omega}$ denotes the free parameters, including $\alpha_{BM}$ and $\Delta_{BM}$, to be determined in the fit. 
The angular parameters of the production process $e^+e^-\to\Lambda^+_c\bar{\Lambda}^-_c$, namely $\alpha_0$ and $\Delta\Phi$, as well as the decay parameters of the subsequent sub-decays, $\alpha_{p\pi^-(\bar{p}\pi^+)}$ and $\alpha_{p\pi^0(\bar{p}\pi^0)}$, are fixed during the fitting procedure.
$\mathcal{P}^{\Lambda^+_c}(\boldsymbol{\xi}_{i};\boldsymbol{\omega})$ is the total PDF of the entire decay chain, and $\epsilon(\boldsymbol{\xi}_{i})$ represents the detection efficiency, parameterized as a function of the kinematic variables $\boldsymbol{\xi}_{i}$. 
The normalization factor is evaluated by integrating over a large phase space MC sample, as given by
\begin{equation}
	\label{integration}
\int\epsilon(\boldsymbol{\xi}_{i})|\mathcal{P}^{\Lambda^+_c}(\boldsymbol{\xi}_{i};\boldsymbol{\omega})|^{2}\text{d}\boldsymbol{\xi}_{i}\propto\frac{1}{N_{\text{gen}}}\sum_{k_{\text{MC}}}^{N_{\text{MC}}}|\mathcal{P}^{\Lambda^+_c}(\boldsymbol{\xi}_{k_{\text{MC}}};\boldsymbol{\omega})|^{2}~,
\end{equation}
where $N_{\text{gen}}$ is the total number of the generated phase space MC events, $N_{\text{MC}}$ is the number of MC events that survive all selection criteria, and $k_{\text{MC}}$ denotes the index of selected MC events.

Using the MINUIT package~\cite{James:1975dr}, we minimize the negative logarithmic likelihood to extract the parameters related to $P$ and $C\!P$ violation. The absolute uncertainties of the $C\!P$-violating parameters are evaluated based on the relative uncertainties of the $P$-violating parameters.

\section{Results}\label{sec:results}
In this section, we present the estimation results on the fitting parameters at STCF and make comparions to the related studies at LHCb and Belle (II). Furthermore, we compare the accuracy of the same parameters in STCF under different polarization of beam and statistics.
The LHCb experiment has a natural advantage in studying the $P$ and $C\!P$-violating parameters of charmed baryons, as the decay of a large number of $\Lambda^0_b$ brings almost 100\% longitudinal polarization to $\Lambda^+_c$, greatly improving the measurement accuracy of the $P$ and $C\!P$-violating parameters. Until now, the LHCb experiment have collected the integral luminosity of $9.0~\text{fb}^{-1}$ $pp$ colission data, and currently a new data collection round is underway. Compared to the previous round of data collection, LHCb experiment have achieved higher detection efficiency.
For the Belle (II) experiment, they have measured many $P$-violating parameters and $C\!P$ asymmetries~\cite{Belle:2022uod}. And there is also a high luminosity data taken plan, which aims to obtain approximately $50~\text{ab}^{-1}$ $e^+e^-$ collision data in 2043~\cite{SuperKEKB2025}.

In our estimation, we take into account the current data-taking conditions and project the future measurement accuracy based on the published data-taking plan~\cite{SuperKEKB2025, LHCb:2012doh}, as illustrated in Fig.~\ref{fig:timeline}.
For the LHCb experiment, the current uncertainty is directly taken from Ref.~\cite{LHCb:2024tnq}, which only uses $\Lambda_c^+$ decays originating from $\Lambda_b^0$.
The STCF faces high integrated luminosity measurements of $300~\text{fb}^{-1}$ for LHCb and $50~\text{ab}^{-1}$ for Belle (II) during operation, which directly affects whether $P$ and $C\!P$-violating measurements for charmed baryon in STCF could be competitive.

\begin{figure*}[htbp]
	\centering
    \subfigure[Longitudinal polarization only.]{\includegraphics[width=0.36\linewidth]{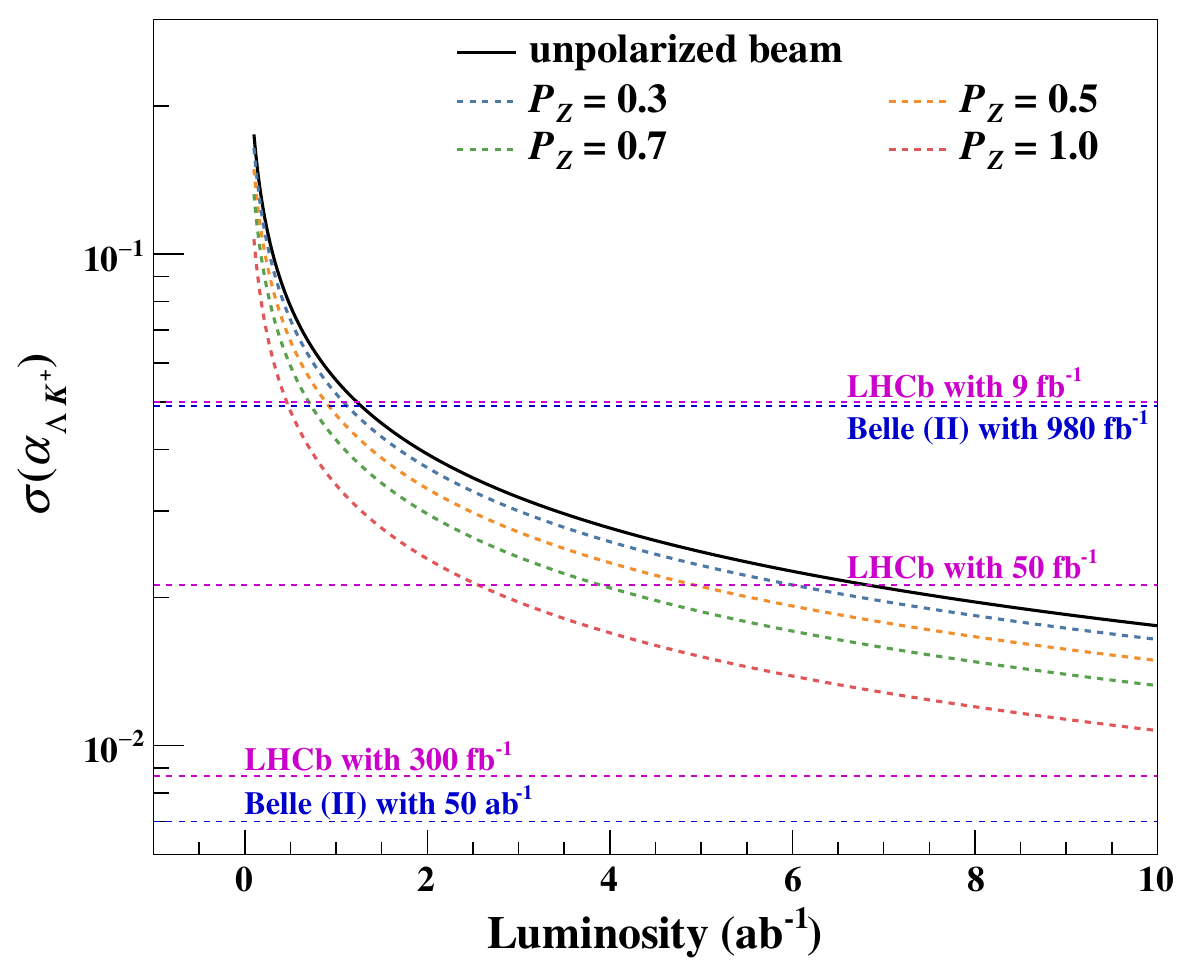}\label{fig:alpha30pz}}
    \quad\quad
	\subfigure[Transverse polarization only.]{\includegraphics[width=0.36\linewidth]{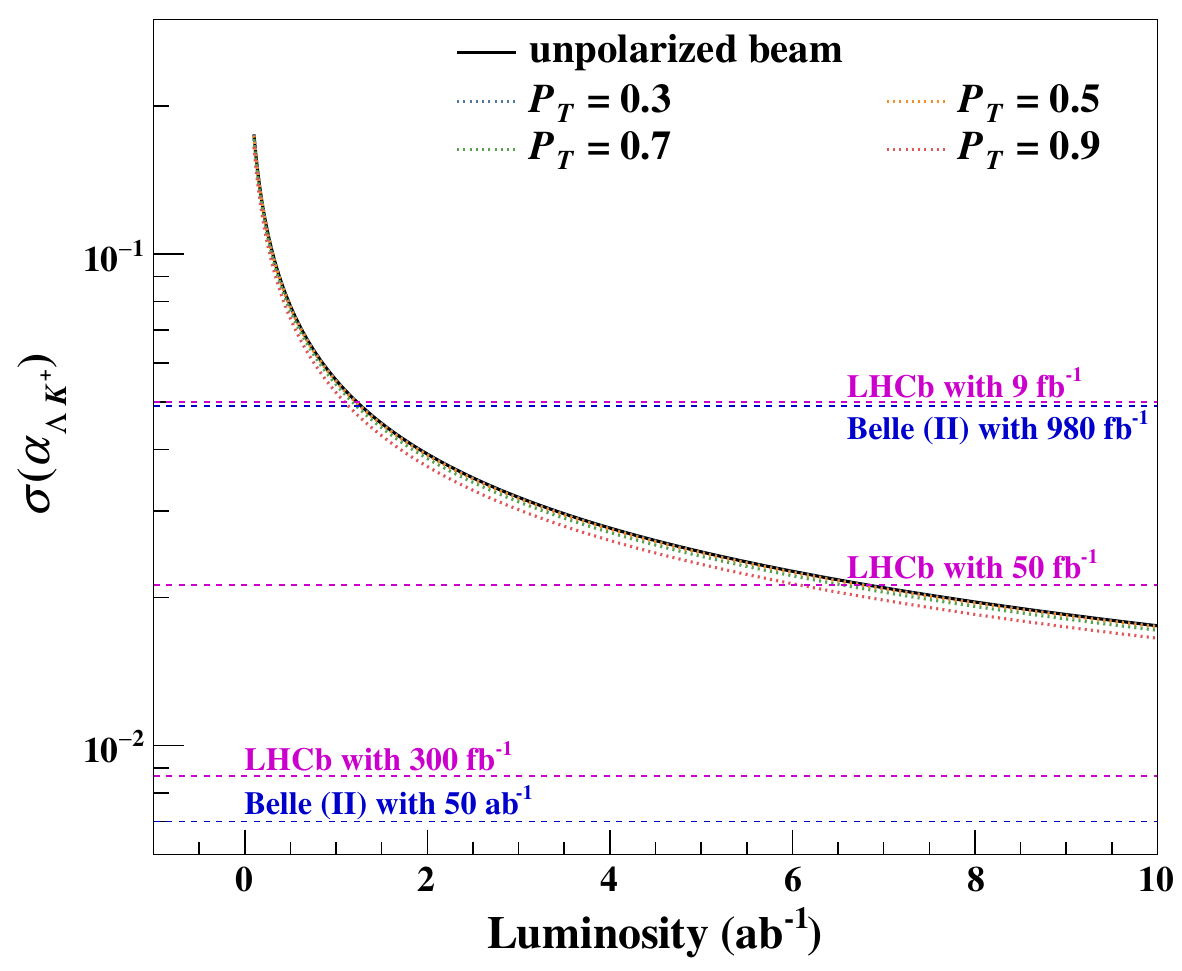}\label{fig:alpha30pt}}
	\caption{Uncertainty prediction of $P$-violating parameters $\alpha_{\Lambda K^+}$ for channel $\Lambda_c^+\to\Lambda K^+$ in different integrated luminosity with different polarized beam.}
	\label{fig:alpha30}
\end{figure*}

\begin{figure*}[htbp]
	\centering
    \subfigure[Uncertainty prediction of $\Delta_{\Lambda K^+}$ with only longitudinal polarization.]{\includegraphics[width=0.36\linewidth]{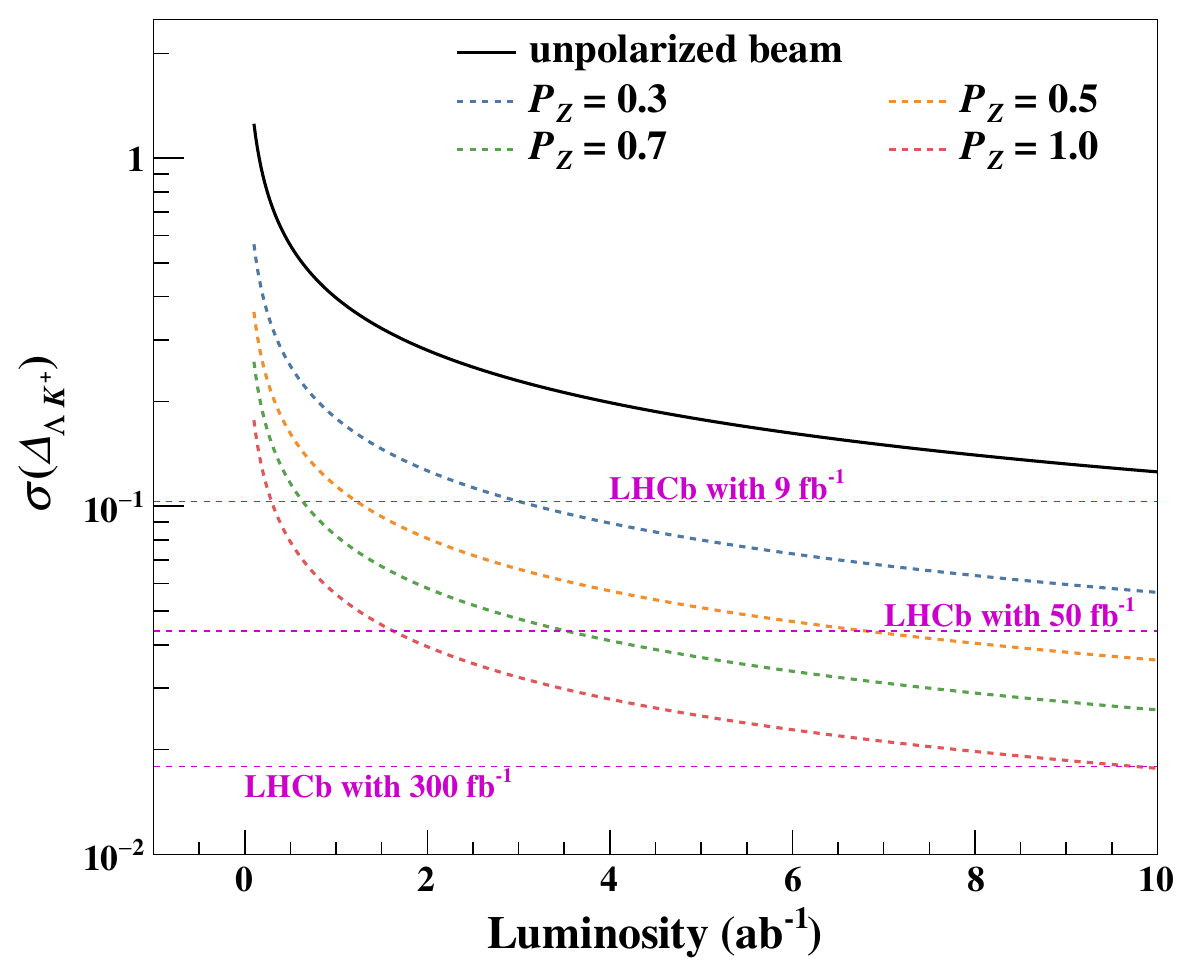}\label{fig:delta30pz}}
    \quad\quad
	\subfigure[Uncertainty prediction of $A^{\alpha_{\Lambda K^+}}_{C\!P}$ with only longitudinal polarization.]{\includegraphics[width=0.36\linewidth]{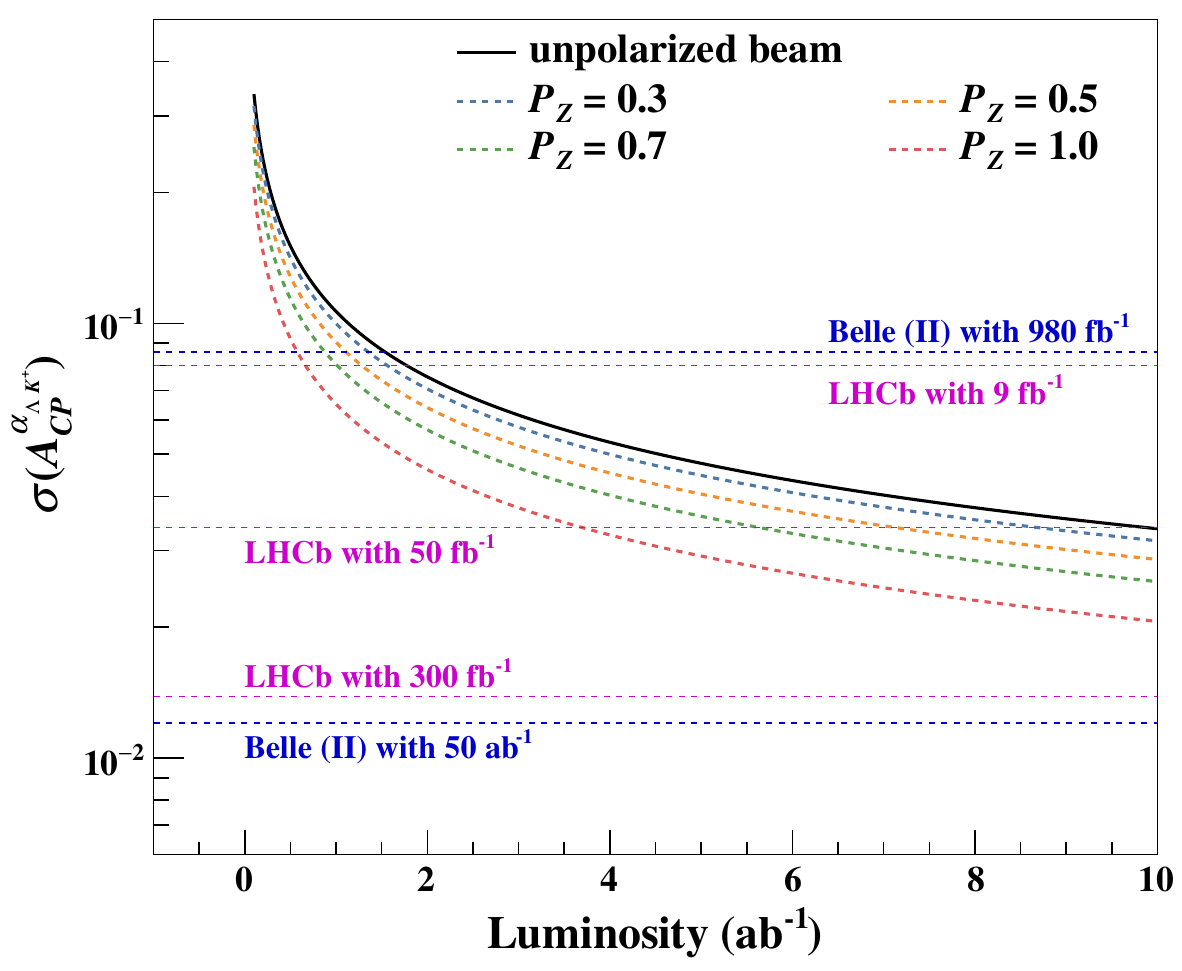}\label{fig:cpv30pz}}
	\caption{Uncertainty prediction of parameters $\Delta_{\Lambda K^+}$ and $A^{\alpha_{\Lambda K^+}}_{C\!P}$ for channel $\Lambda_c^+\to\Lambda K^+$ in different integrated luminosity with different polarized beam.}
	\label{fig:other30}
\end{figure*}

\begin{figure*}[htbp]
	\centering
    \subfigure[Uncertainty prediction of $\alpha_{p\pi^0}$ with only longitudinal polarization.]{\includegraphics[width=0.36\linewidth]{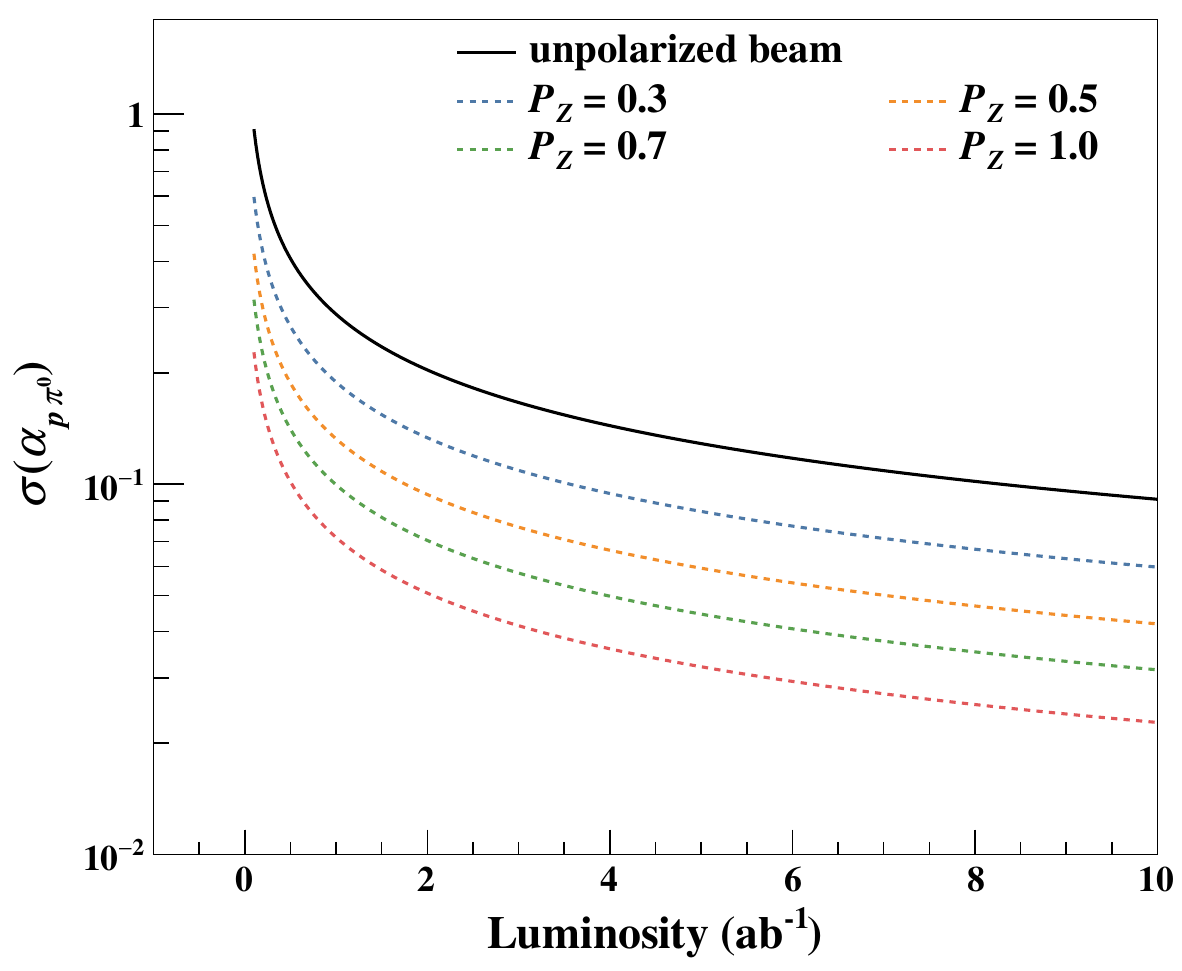}\label{fig:alpha70pz}}
    \quad\quad
	\subfigure[Uncertainty prediction of $A^{\alpha_{p\pi^0}}_{C\!P}$ with only longitudinal polarization.]{\includegraphics[width=0.36\linewidth]{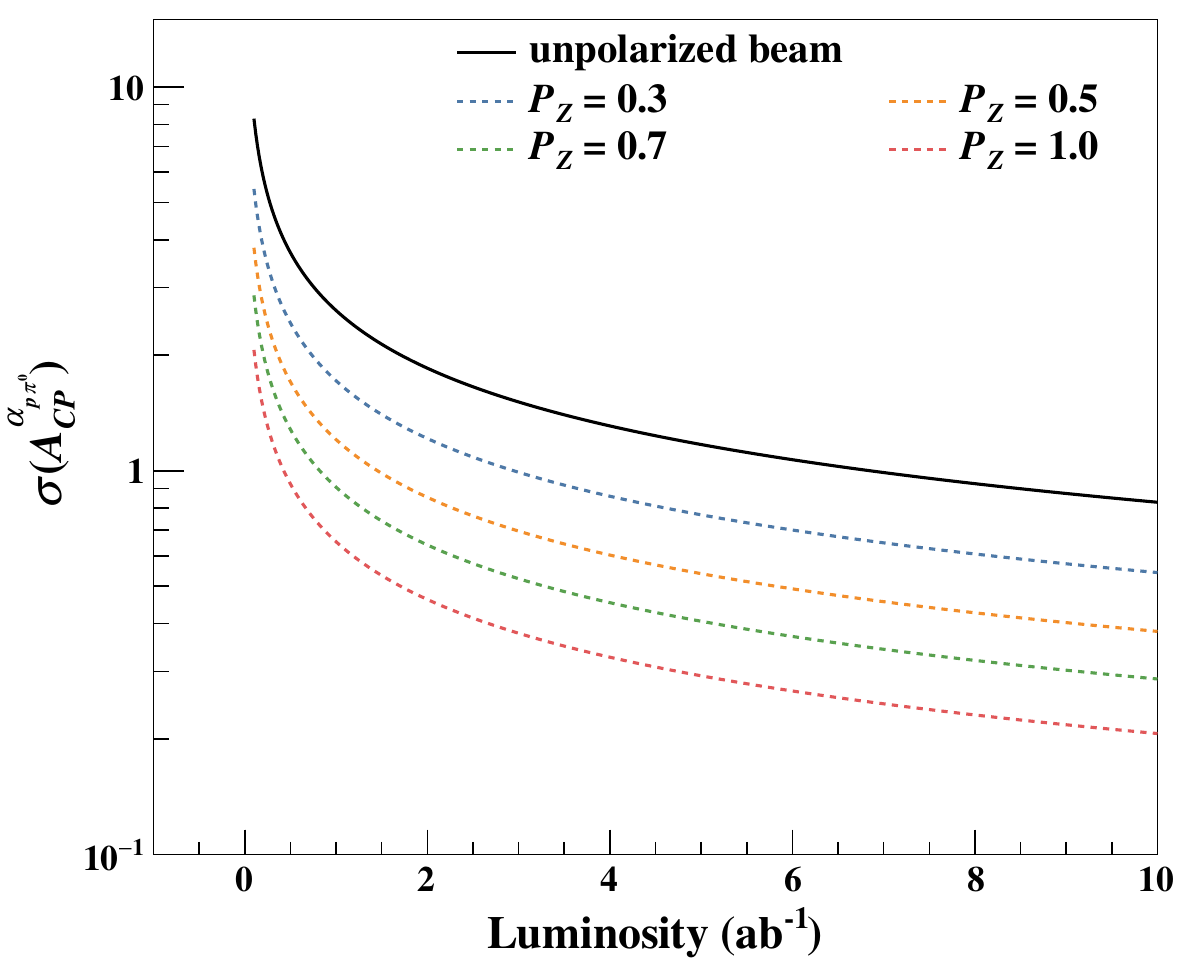}\label{fig:cpv70pz}}
	\caption{Uncertainty prediction of parameters $\alpha_{p\pi^0}$ and $A^{\alpha_{p\pi^0}}_{C\!P}$ for channel $\Lambda_c^+\to p\pi^0$ in different integrated luminosity with different polarized beam.}
	\label{fig:70}
\end{figure*}

\begin{figure*}[htbp]
	\centering
    \subfigure[Uncertainty prediction of $\alpha_{p\eta}$ with only longitudinal polarization.]{\includegraphics[width=0.36\linewidth]{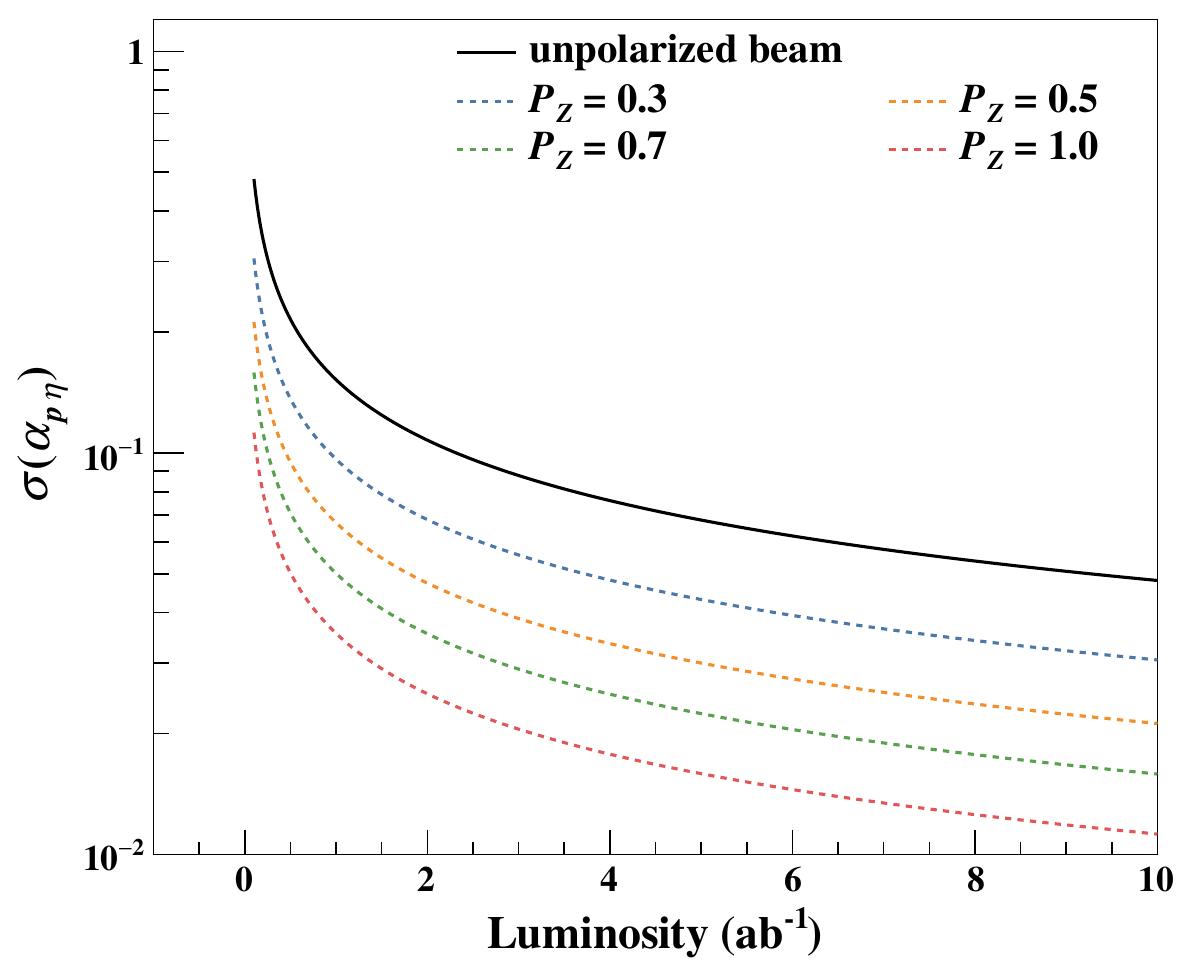}\label{fig:alpha0pz}}
    \quad\quad
	\subfigure[Uncertainty prediction of $A^{\alpha_{p\eta}}_{C\!P}$ with only longitudinal polarization.]{\includegraphics[width=0.36\linewidth]{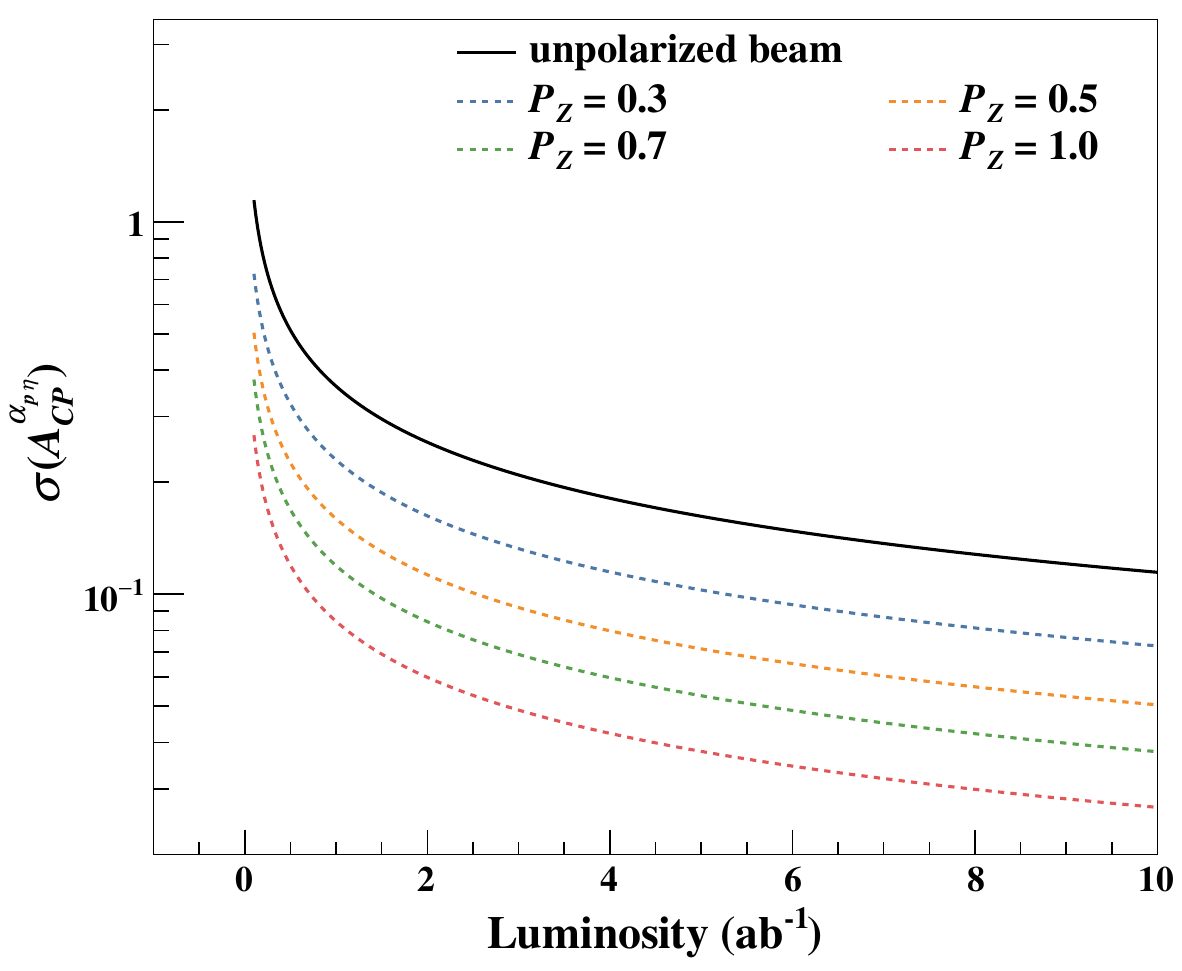}\label{fig:cpv0pz}}
	\caption{Uncertainty prediction of parameters $\alpha_{p\eta}$ and $A^{\alpha_{p\eta}}_{C\!P}$ for channel $\Lambda_c^+\to p\eta$ in different integrated luminosity with different polarized beam.}
	\label{fig:0}
\end{figure*}

\begin{figure*}[htbp]
	\centering
    \subfigure[Uncertainty prediction of $\alpha_{\Sigma^0 K^+}$ with only longitudinal polarization.]{\includegraphics[width=0.3\linewidth]{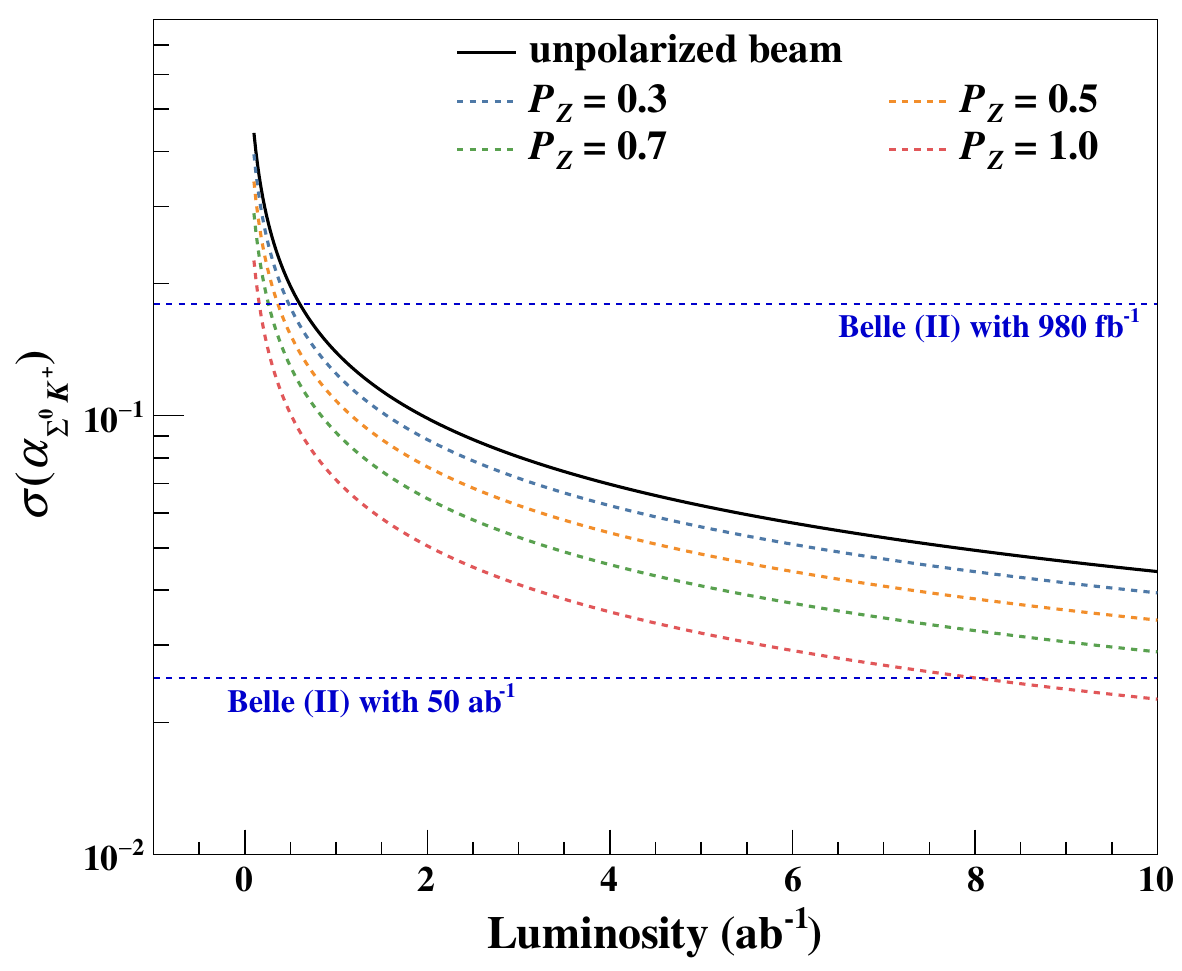}\label{fig:alpha60pz}}
    \quad\quad
    \subfigure[Uncertainty prediction of $\Delta_{\Sigma^0 K^+}$ with only longitudinal polarization.]{\includegraphics[width=0.3\linewidth]{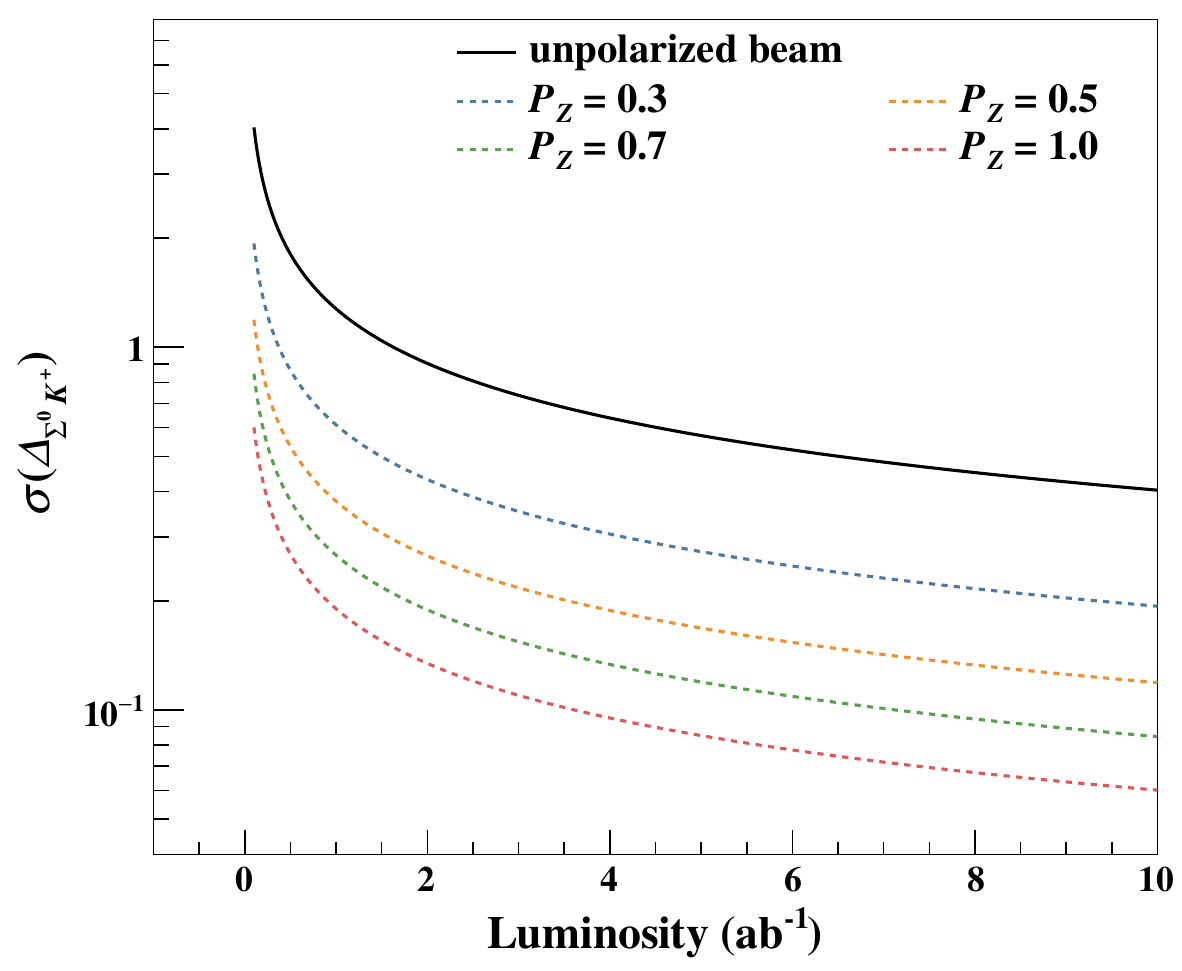}\label{fig:delta60pz}}
    \quad\quad
	\subfigure[Uncertainty prediction of $A^{\alpha_{\Sigma^0 K^+}}_{C\!P}$ with only longitudinal polarization.]{\includegraphics[width=0.3\linewidth]{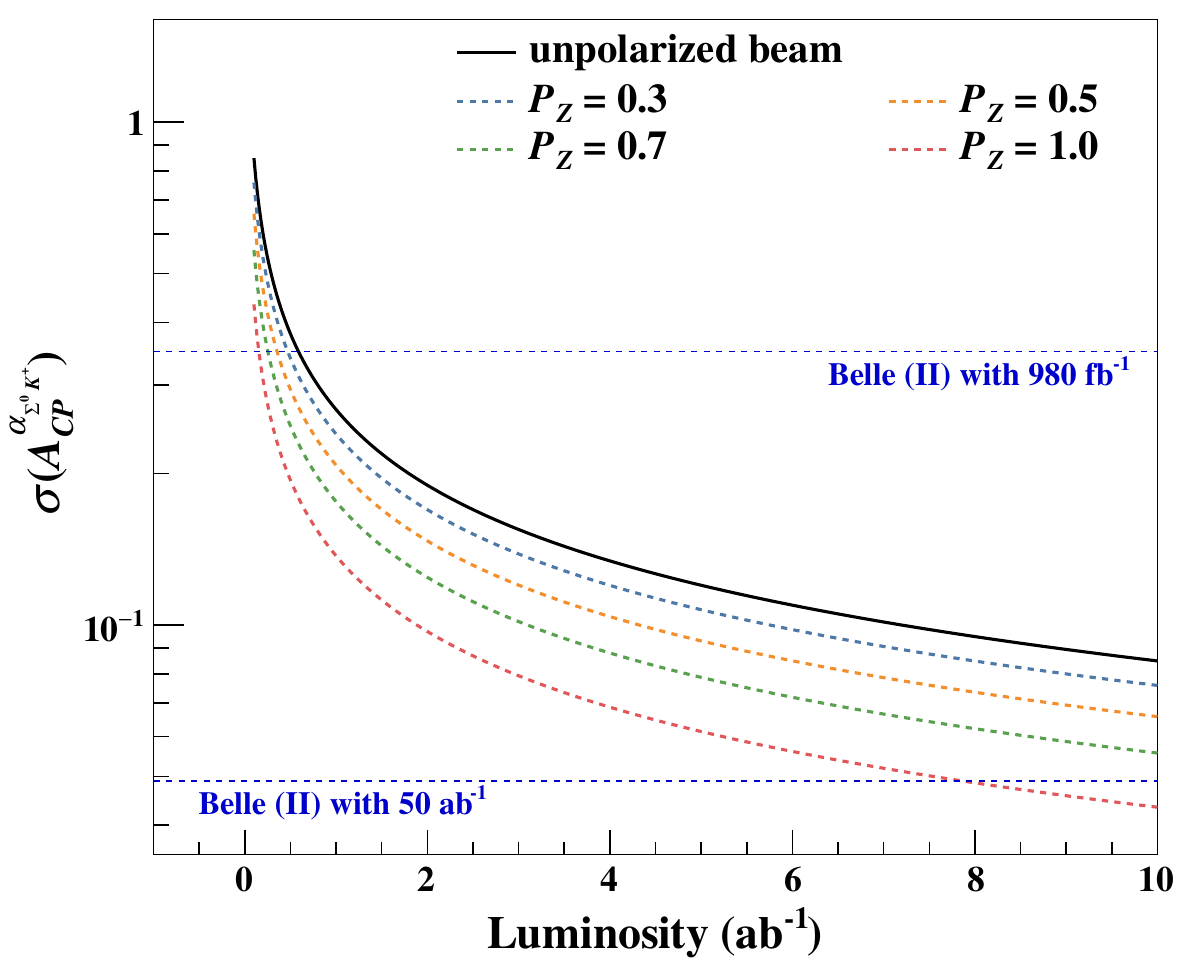}\label{fig:cpv60pz}}
	\caption{Uncertainty prediction of parameters $\alpha_{\Sigma^0 K^+}$, $\Delta_{\Sigma^0 K^+}$, and $A^{\alpha_{\Sigma^0 K^+}}_{C\!P}$ for channel $\Lambda_c^+\to\Sigma^0 K^+$ in different integrated luminosity with different polarized beam.}
	\label{fig:60}
\end{figure*}

\begin{figure*}[htbp]
	\centering
    \subfigure[Uncertainty prediction of $\alpha_{\Sigma^+ K^0_S}$ with only longitudinal polarization.]{\includegraphics[width=0.3\linewidth]{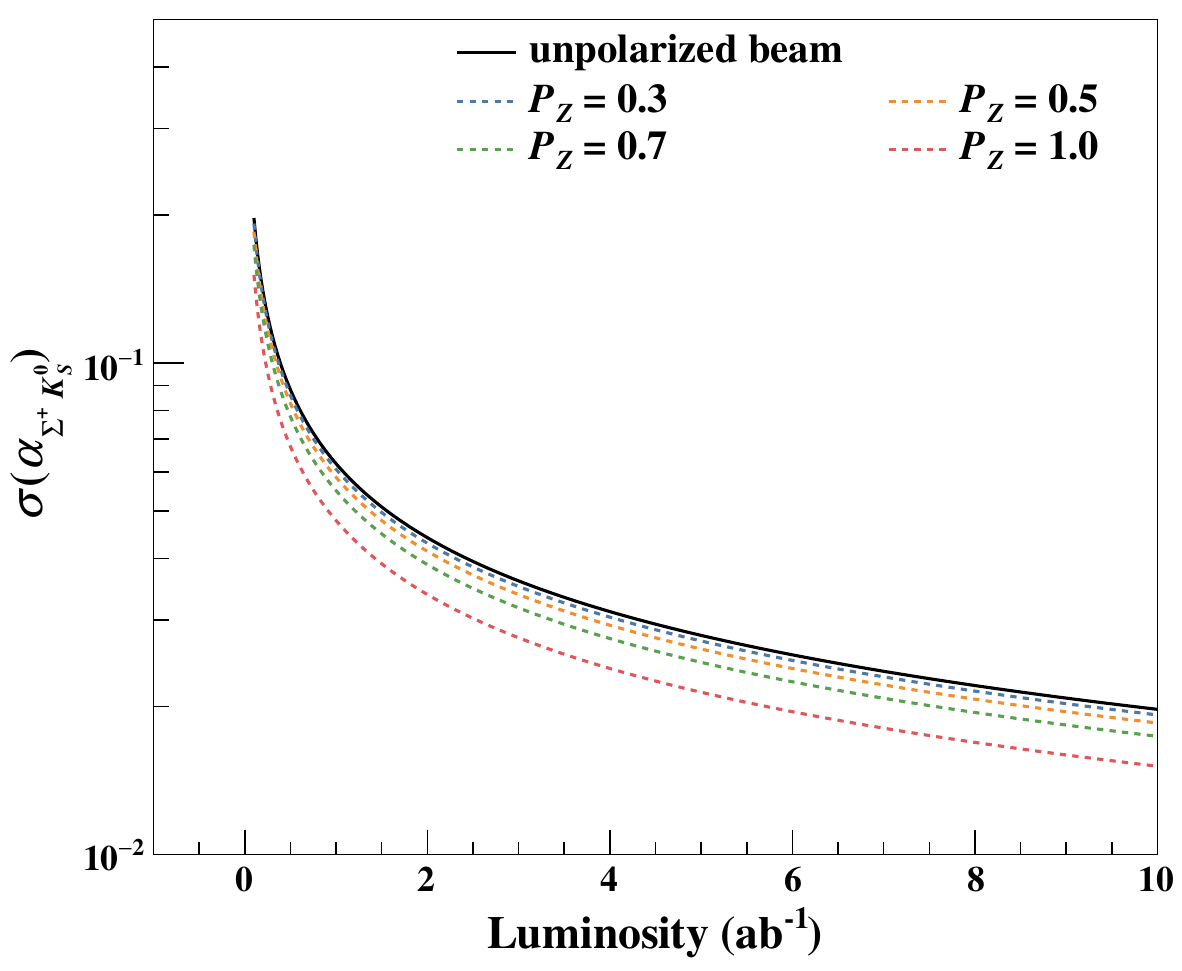}\label{fig:alpha62pz}}
    \quad\quad
    \subfigure[Uncertainty prediction of $\Delta_{\Sigma^+ K^0_S}$ with only longitudinal polarization.]{\includegraphics[width=0.3\linewidth]{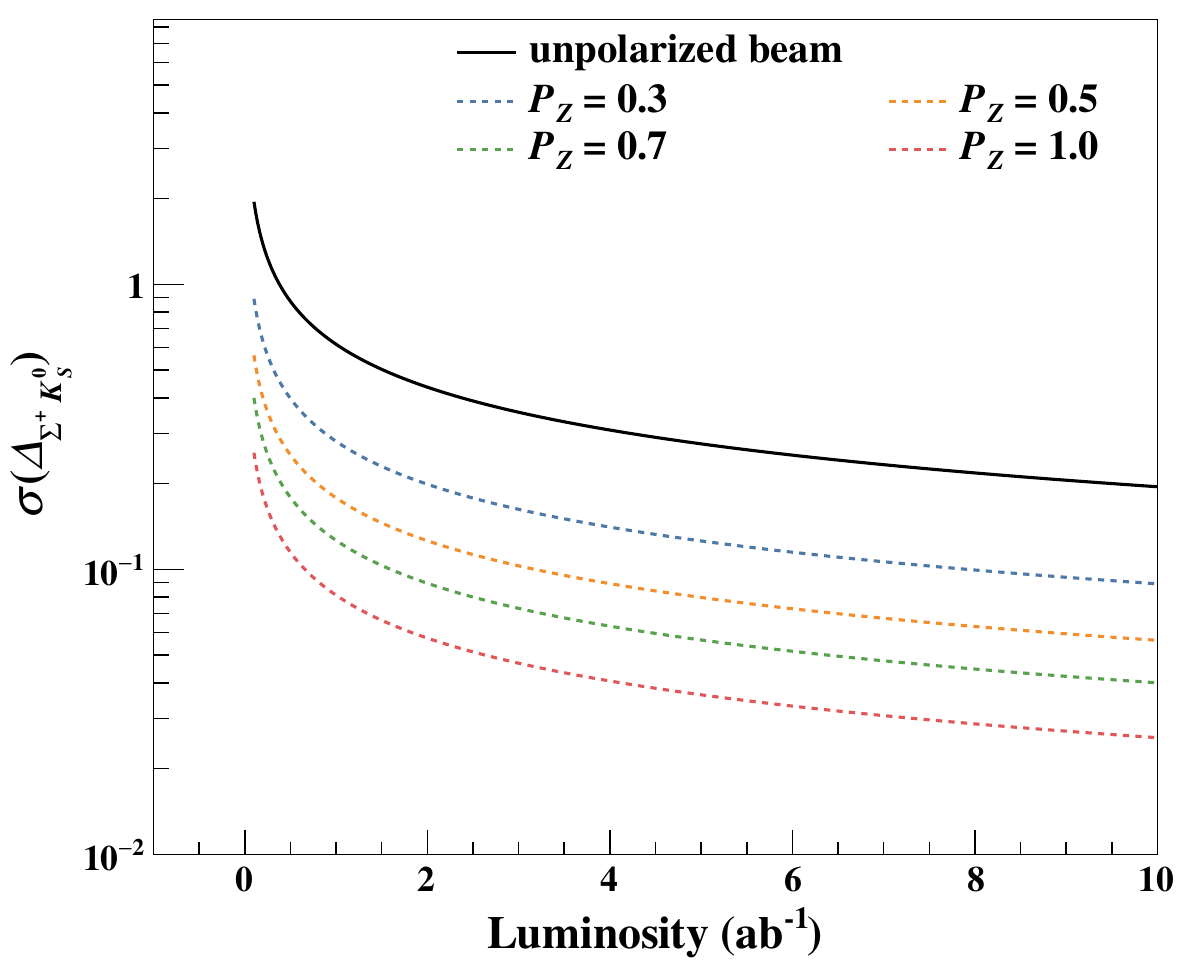}\label{fig:delta62pz}}
    \quad\quad
	\subfigure[Uncertainty prediction of $A^{\alpha_{\Sigma^+ K^0_S}}_{C\!P}$ with only longitudinal polarization.]{\includegraphics[width=0.3\linewidth]{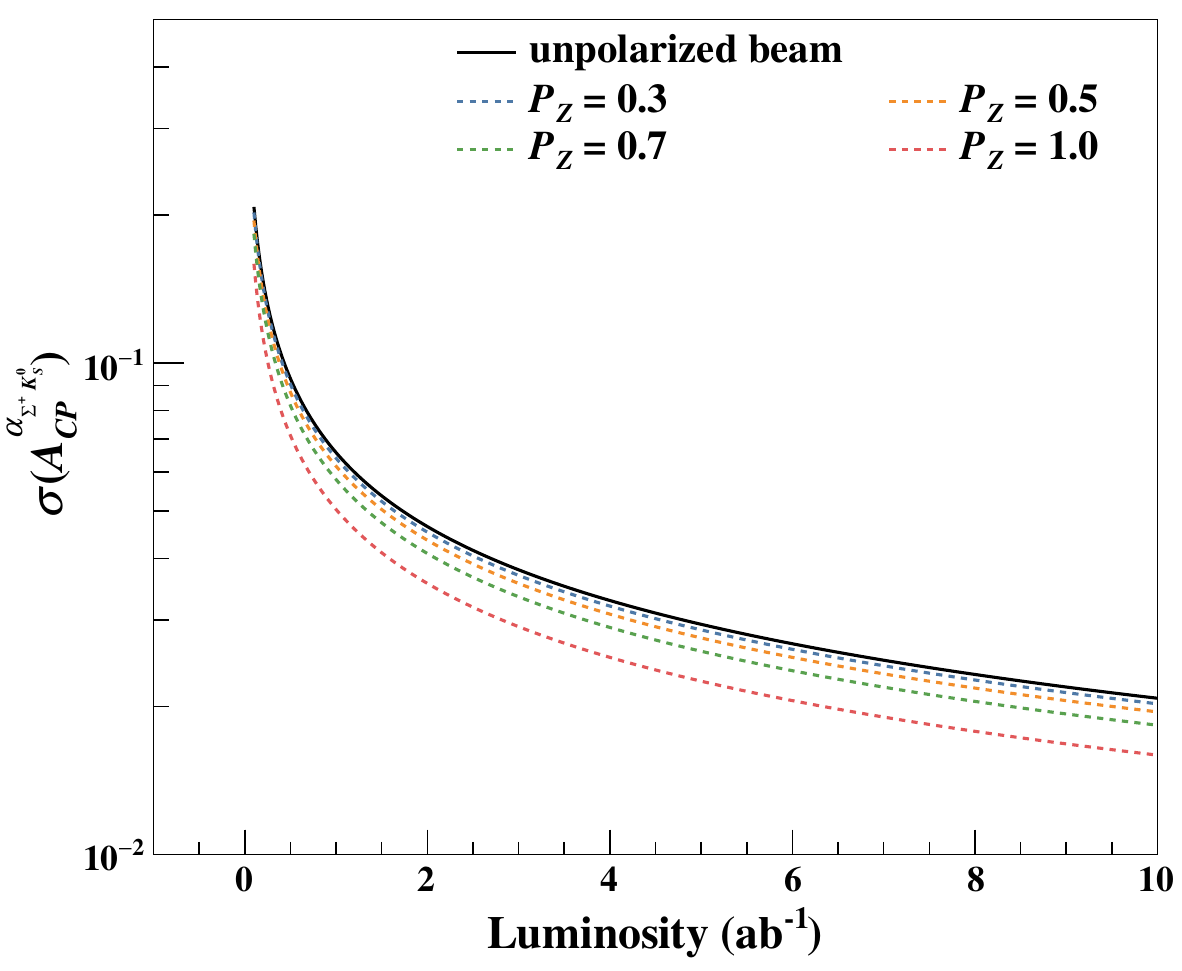}\label{fig:cpv62pz}}
	\caption{Uncertainty prediction of parameters $\alpha_{\Sigma^+ K^0_S}$, $\Delta_{\Sigma^+ K^0_S}$, and $A^{\alpha_{\Sigma^+ K^0_S}}_{C\!P}$ for channel $\Lambda_c^+\to\Sigma^+ K^0_S$ in different integrated luminosity with different polarized beam.}
	\label{fig:62}
\end{figure*}

After performing the fit procedure in Section~\ref{met:fit}, the accuracy of $P$ and $C\!P$-violating parameters is obtained under different beam polarizations, as shown in Fig.~\ref{fig:alpha30}, Fig.~\ref{fig:other30}, Fig.~\ref{fig:70}, Fig.~\ref{fig:0}, Fig.~\ref{fig:60}, and Fig.~\ref{fig:62}.
The current and future accuracies in the LHCb and Belle (II) experiments listed in Tab.~\ref{future} are overlaid in the plots.
As shown in Fig.~\ref{fig:alpha30}, for the decay $\Lambda^+_c\to\Lambda K^+$, the precision of the $P$-violating parameter $\alpha_{\Lambda K^+}$ benefits significantly from longitudinal beam polarization, while the improvement induced by transverse polarization is marginal. The sensitivity increases with increasing longitudinal polarization $P_Z$, and with $P_Z = 1.0$, the statistical uncertainty can approach $10^{-2}$ with a $10~\mathrm{ab}^{-1}$ of integrated luminosity. 
Figure~\ref{fig:other30} presents the uncertainty projections for $\Delta_{\Lambda K^+}$ and $A_{C\!P}^{\alpha_{\Lambda K^+}}$ in the same decay. Interestingly, the gain in precision for $\Delta_{\Lambda K^+}$ shows diminishing returns as $P_Z$ increases, contrary to the trend seen in $\alpha_{\Lambda K^+}$.

\begin{table}[htbp]
    \small
	\centering
	\caption{Current and projected accuracies at LHCb and Belle (II) with the corresponding luminosities, where absolute values of the uncertainties are given. The quoted luminosities in nowadays are those used in the latest measurements~\cite{LHCb:2024tnq,Belle:2022uod}.}
	\label{future}
	\begin{tabular}{c c c c c c}
		\hline\hline
		\multirow{2}*{Parameters}           & \multicolumn{3}{c}{LHCb}          & \multicolumn{2}{c}{Belle (II)}  \\
        \cline{2-6}
        ~                                   & Nowadays      & 2030         & 2041          & Nowadays                         & 2043  \\
		\hline
        Luminosity                          & 9~fb$^{-1}$   & 50~fb$^{-1}$ & 300~fb$^{-1}$ & 980~fb$^{-1}$ & 50~ab$^{-1}$ \\  
        \hline
        $\alpha_{\Lambda K^+}$              & 0.05          & 0.021        & 0.009         & 0.049         & 0.007 \\
        $\Delta_{\Lambda K^+}$              & 0.103         & 0.044        & 0.018         & $\cdots$      & $\cdots$ \\
        $\alpha_{\Sigma^0 K^+}$             & $\cdots$      & $\cdots$     & $\cdots$      & 0.18          & 0.025 \\
        $A^{\alpha_{\Lambda K^+}}_{C\!P}$    & 0.08          & 0.034        & 0.014         & 0.086         & 0.012 \\
        $A^{\alpha_{\Sigma^0 K^+}}_{C\!P}$   & $\cdots$      & $\cdots$     & $\cdots$      & 0.35          & 0.049 \\
        \hline\hline
	\end{tabular}
\end{table}

Figures~\ref{fig:70} and \ref{fig:0} show the projected uncertainties for $\alpha_{p\eta}$, $A_{C\!P}^{\alpha_{p\eta}}$, and $\alpha_{p\pi^0}$, $A_{C\!P}^{\alpha_{p\pi^0}}$ in the $\Lambda_c^+\to p\eta$ and $\Lambda_c^+\to p\pi^0$ decay, respectively. These final states include photons, which is challenging for LHCb due to limited photon detection efficiency. The STCF, equipped with excellent photon reconstruction capability, can reach sensitivities at the $10^{-2}$ level with a polarized beam and $10~\text{ab}^{-1}$ luminosity. This gives STCF a unique and irreplaceable role in probing symmetry-violating observables in such decays.

In Fig.~\ref{fig:60}, the predicted sensitivities for $\alpha_{\Sigma^0 K^+}$ and $A_{C\!P}^{\alpha_{\Sigma^0 K^+}}$ demonstrate that STCF with a large $P_Z$ clearly outperforms Belle (II), especially at high luminosity. Due to the presence of photons from $\Sigma^0 \to \Lambda \gamma$, this mode is challenging for LHCb, underscoring the importance of STCF and Belle (II) in accessing this channel.
Fig.~\ref{fig:62} illustrates the sensitivity to $\alpha_{\Sigma^+ K_S^0}$, $\Delta_{\Sigma^+ K_S^0}$, and $A_{C\!P}^{\alpha_{\Sigma^+ K_S^0}}$. For all three observables, the precision improves as $P_Z$ increases.

\section{Discussion}\label{sec:dis}
In this study, we examine the enhancement of measurement precision for $P$ and $C\!P$-violating parameters in the next-generation $e^+e^-$ collider STCF, incorporating transverse and longitudinal beam polarization.
Using multi-dimensional fits to angular distributions, we evaluate the expected uncertainties of polarization parameters in the weak decays $\Lambda^+_c\to p\pi^0$, $p\eta$, $\Lambda K^+$, $\Sigma^0 K^+$, and $\Sigma^+ K^0_S$.
The results reveal that longitudinal polarization significantly improves the experimental accuracy compared to the transverse polarization, as the comparisons of the uncertainties between in Fig.~\ref{fig:alpha30pz} and Fig.~\ref{fig:alpha30pt}.

To systematically assess the sensitivity of $P$ and $C\!P$ violation measurements, we select decay modes featuring different decay topologies and different final-state baryons such as $p$, $\Lambda$, $\Sigma^{+}$, and $\Sigma^{0}$.
These differences result in distinct angular distributions and detection efficiencies, thereby offering varying sensitivities to the underlying physical parameters.

With the projected data statistics from one year of STCF operation, especially under conditions of high longitudinal polarization, the expected precision on parameters such as $\Delta_{\Lambda K^+}$, $\alpha_{\Sigma^0 K^+}$, and $A^{\alpha_{\Sigma^0 K^+}}_{C\!P}$ is anticipated to exceed that achievable at other current or planned experiments in the future.
Specifically, the parameter $\Delta$ exhibits enhanced sensitivity to beam polarization, which in turn leads to a significantly improved precision in the determination of $\beta$.
Consequently, STCF has a unique capability to probe the $A^{\beta}_{C\!P}$ parameter, thereby filling a critical gap in $C\!P$ violation studies.

\begin{figure}[htbp]
	\centering
        \includegraphics[width=0.95\linewidth]{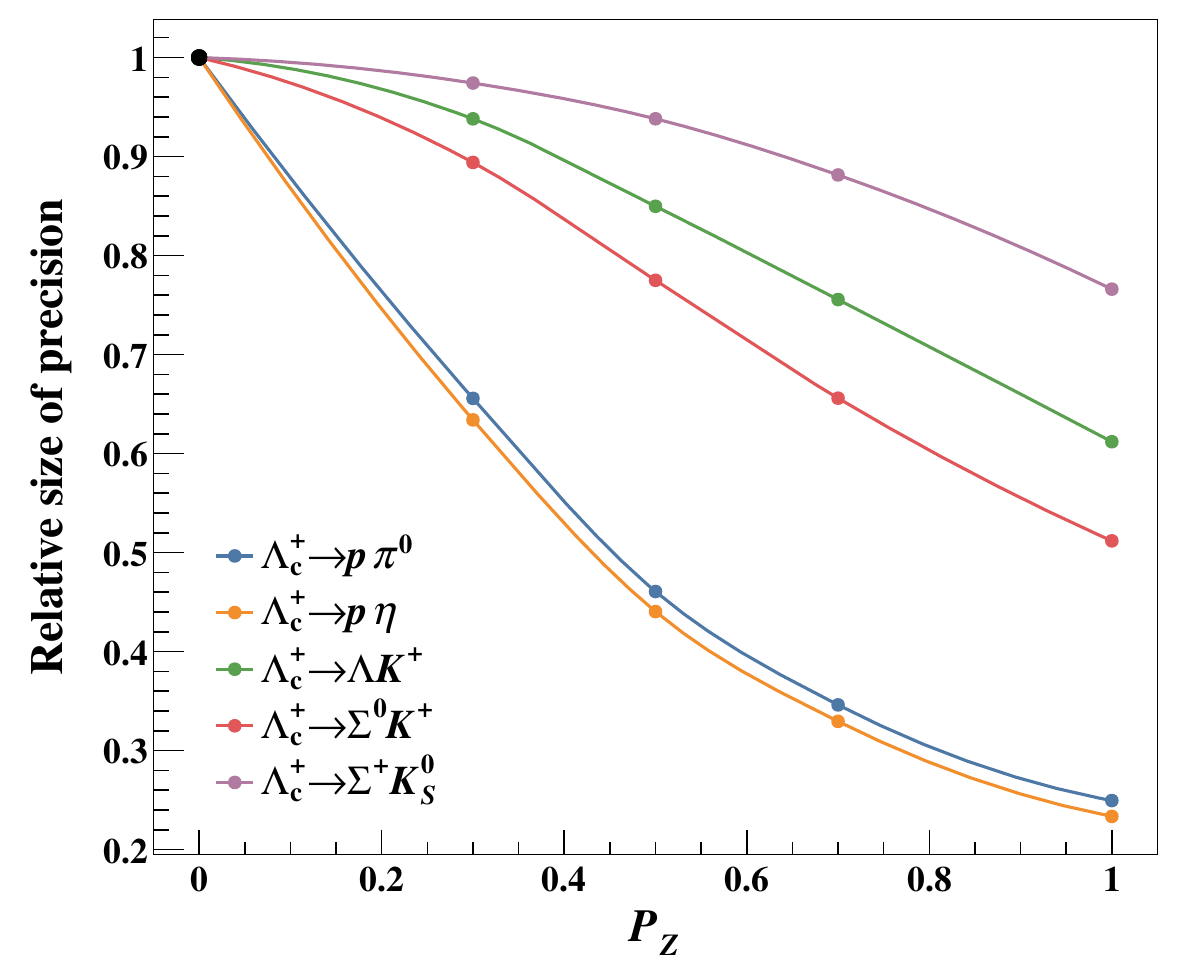}
	\caption{Relative size of $\alpha$ and $A^{\alpha}_{C\!P}$ precision at different $P_Z$ values with respect to the unpolarized case for a 1 ab$^{-1}$ dataset.}
	\label{fig:PZimprovement}
\end{figure}

In conclusion, our study highlights the substantial impact of beam polarization on the precision of $P$ and $C\!P$ violation measurements in charmed baryon decays.
As shown in Fig.~\ref{fig:PZimprovement}, longitudinal beam polarization is crucial for measuring $P$ and $C\!P$ violation, and with about 70\% polarization, the achievable precision can be improved by up to a factor of three, depending on the decay mode.
The STCF, with its ability to provide highly polarized beams and accumulate large data samples, stands out as a powerful platform for exploring subtle symmetry-breaking effects.
Looking ahead, STCF is expected to deliver competitive precision in these measurements and maintain significant advantages over existing and planned facilities.

\section*{Acknowledgement}
This work is supported in part by the National Key R\&D Program of China under Contract No. 2022YFA1602200; 
International partnership program of the Chinese Academy of Sciences Grant No. 211134KYSB20200057; 
National Natural Science Foundation of China (NSFC) under Contracts Nos. 12422504, 12221005 and 12175244;
Fundamental Research Funds for the Central Universities, Lanzhou University under Contracts Nos. lzujbky-2025-ytB01, lzujbky-2023-stlt01, lzujbky-2023-it32, and University of Chinese Academy of Sciences.
We thank the Hefei Comprehensive National Science Center for their strong support on the STCF key technology research project.

\bibliographystyle{apsrev4-1}
\bibliography{bibliography}

\end{document}